\documentclass[]{fairmeta}

\title{Genetic Algorithm-Accelerated Computational Discovery of Liquid Crystal Polymers with Enhanced Optical Properties}

\usepackage{xcolor}
\usepackage{graphicx}
\usepackage{cite}
\usepackage{natbib}
\usepackage{multirow}
\usepackage{amsmath}
\usepackage{booktabs}

\author[1,*]{Jianing Zhou}
\author[1,+]{Yuge Huang}
\author[1]{Arman Boromand}
\author[1]{Keian Noori}
\author[1]{Lafe Purvis}
\author[1]{Chulwoo Oh}
\author[1]{Lu Lu}
\author[2]{Zachary W. Ulissi}
\author[2,\dagger]{Vahe Gharakhanyan}
\author[1,\dagger]{Xinyue Zhang}

\affiliation[1]{Meta Reality Labs}
\affiliation[2]{Meta Fundamental AI Research}

\contribution[*]{Work done during Meta internship}
\contribution[+]{Work done while at Meta}
\contribution[\dagger]{Joint last author}

\abstract{Liquid crystal polymers with exceptional optical properties are highly promising for next-generation virtual, augmented, and mixed reality (VR/AR/MR) technologies, serving as high-performance, compact, lightweight, and cost-effective optical components. However, the growing demands for optical transparency and high refractive index in advanced optical devices present a challenge for material discovery. In this study, we develop a novel approach that integrates first-principles calculations with genetic algorithms to accelerate the discovery of liquid crystal polymers with low visible absorption and high refractive index. By iterating within a predefined space of molecular building blocks, our approach rapidly identifies reactive mesogens that meet target specifications. Additionally, it provides valuable insights into the relationships between molecular structure and properties. This strategy not only accelerates material screening but also uncovers key molecular design principles, offering a systematic and scalable alternative to traditional trial-and-error methods.}

\date{\today}
\correspondence{Vahe Gharakhanyan \email{vaheg@meta.com}, Xinyue Zhang \email{xinyuezhang@meta.com}}

\begin{document}

\maketitle

\section{Introduction}
\label{section:intro}

Liquid crystal polymers (LCPs) have played an important role in the development of advanced optical components. By leveraging the optical anisotropy of the liquid crystal molecules and their flexible alignment capabilities, liquid crystal polymer enable many thin-film based planar optical elements that are high-efficiency, compact, lightweight, and cost-effective, with examples including photochromic films, optical fibers, and polarization holographic gratings and lenses.~\citep{shibaev2003photoactive, li2016fabrication, chen2020liquid} With the rapid advancement of display technologies and virtual, augmented, and mixed reality (VR/AR/MR) systems, liquid crystal polymer-based spatial light modulators and waveguides have attracted significant interest due to their critical functionality in these devices.~\citep{kim2015fabrication, lee2017reflective, xiong2021planar} 

A key advantage of liquid crystal polymers in optical applications is their combination of low optical absorption in the visible range and high refractive indices, which improves light efficiency while minimizing component weight and volume.~\citep{zhang202483} The fundamental building blocks of liquid crystal polymers, known as reactive mesogens (RMs), consist of a rigid mesogenic core, alkyl chain spacers, and polymerizable terminal groups, as depicted in \autoref{fig:RM_LCP}(a). The mesogenic core retains liquid crystal properties due to its rod-like conjugated structure, which results in molecular anisotropy. Strong intermolecular interactions between mesogenic cores promote self-assembly and alignment in the liquid crystal phase. The length of the alkyl spacer can be adjusted to modify the liquid crystal phase transition temperature. Additionally, reactive mesogens act as monomers through polymerizable end groups, allowing polymerization by thermal or UV curing to form stable optical films.~\citep{liu2014liquid} The unique structures of reactive mesogens enable tunability of both optical absorption and refractive index, providing a molecular design strategy for enhancing optical properties.

Over the past few decades, design and synthesis of reactive mesogens with desired optical properties remain a significant challenge due to both the theoretical complexity of molecular design and the inefficiency of material discovery.~\citep{wei2016hyperbranched, liu2009high} Mesogenic cores with longer conjugation generally exhibit increased polarizability, enhancing the refractive index. However, this also tends to increase light absorption at longer wavelengths, as the delocalized electrons are more easily excited by lower energy.~\citep{sasagane1993higher, li2005refractive} The experimental process often relies on empirical methods and modifications of existing molecular structures, making it time-consuming and resource-intensive.~\citep{sekine2001high, he2019fabrication, allen2021synthesis} Although computational methods can accelerate molecular design compared to experimental approaches, they also face challenges. Unlike crystal materials with periodic structures, simulating the optical properties of liquid crystal polymers traditionally requires modeling multi-molecular systems, making it similarly time-consuming and unsuitable for large-scale screening. Variations in intramolecular and intermolecular interactions in liquid crystal polymers can significantly influence their optical properties. Therefore, accurate modeling largely relies on capturing the molecular ordering of the liquid crystal polymer system.~\citep{wilson2007molecular}

To address these challenges, it is crucial to develop a pipeline that enables the rapid and accurate simulation of the optical properties of liquid crystal polymers, specifically focusing on absorption and refractive index. This pipeline would facilitate extensive screening of candidate structures, providing insights into structure-property relationships and opening new avenues in the molecular design space. Recent advancements in optical material exploration, particularly through machine learning and generative models, have provided valuable insights for liquid crystal polymer development.~\citep{stein2019machine, simine2020predicting, chen2020machine} However, a comprehensive liquid crystal material database with detailed optical data is still lacking, underscoring the need to explore the liquid crystal polymer space using advanced search and generative methods.

In this study, we develop a first-principles-based computational framework for large-scale screening of liquid crystal polymers with optical transparency and refractive index. By approximating the polymer network configuration using dimer conformations of reactive mesogens in nematic phase, we significantly reduce the computational cost of multi-molecular system simulations while retaining key molecular interactions. Further validation through molecular dynamics (MD) simulations and experimental data confirm the accuracy of this approach, supporting the dimer-based modeling strategy. We then integrate a genetic algorithm to screen, optimize, and generate novel liquid crystal polymer candidates. The resulting molecules demonstrate low absorbance in the visible wavelengths and exceptionally high refractive indices, offering valuable guidance for molecular design and material discovery. Furthermore, the computational pipeline developed in this research is adaptable to other organic molecule discovery platforms, particularly for polymer materials with specific optical properties. This adaptability holds benefits for a wide range of applications in optics, robotics, and other fields, providing a versatile tool for advancing material design and discovery.

\section{Computational Pipeline Methodology and Validation}
\label{section:Computational Pipeline Methodology and Validation}

To facilitate high-throughput computational molecular screening, it is essential to develop a streamlined model that incorporates necessary approximations, supported by robust scientific evidence and experimental validation. In this section, we first introduce liquid crystal polymer networks. The configuration of these networks is simplified as several dimer conformations, which are then used for absorption and refractive index calculations using time-dependent density functional theory (TD-DFT) and density functional theory (DFT), respectively.

\subsection{Formation of Liquid Crystal Polymer Network}

\begin{figure}
    \centering
    \includegraphics[width=0.5\textwidth]{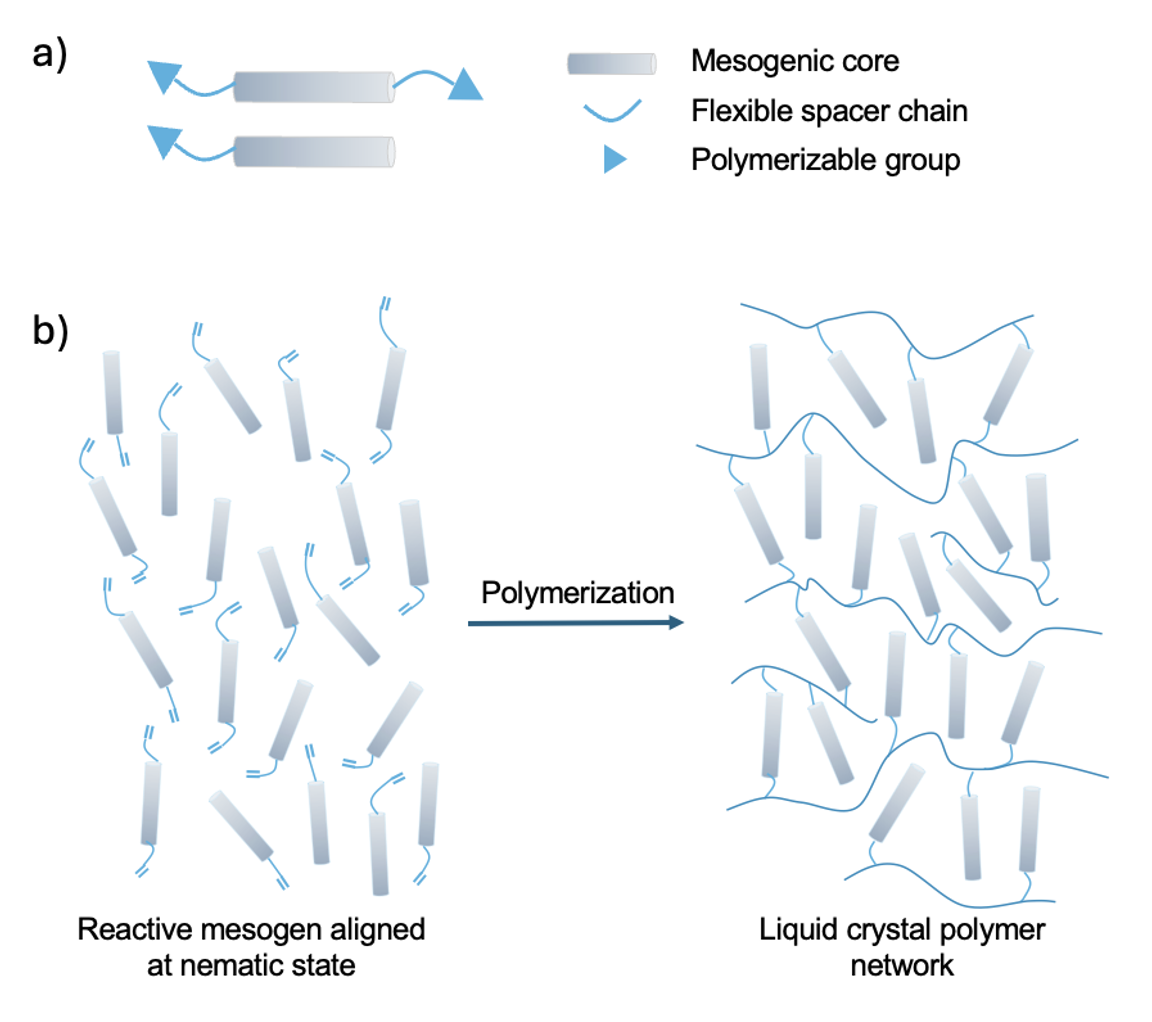}
    \caption{(a) Structures of reactive mesogen. (b) Alignment of reactive mesogens in the nematic phase, followed by polymerization into a liquid crystal polymer network.}
    \label{fig:RM_LCP}
\end{figure}

The formation of a liquid crystal polymer network, depicted in \autoref{fig:RM_LCP}(b), begins by heating reactive mesogens to the liquid crystal phase to achieve alignment. This is followed by a curing step to form a cross-linked network, effectively locking the molecular alignment of reactive mesogens and preserving their conformations in the liquid crystal polymer. Consequently, the optical properties of the liquid crystal polymer are largely determined by the packing of the reactive mesogens.~\citep{shibaev2003photoactive} Although minor differences in the order parameter and density between reactive mesogens and corresponding liquid crystal polymers arise due to spacer shrinkage before and after cross-linking process, these differences have a smaller impact compared to the influence of molecular arrangement on optical properties.~\citep{liu2014liquid, cooper2024controlling} In this work, we use the conformational states of reactive mesogens in the nematic phase as an approximation for the polymer network when simulating the optical properties of liquid crystal polymers, as discussed in the following sections.

\subsubsection{Dimer Simplifications}

The arrangement and packing of reactive mesogens are crucial in determining the optical properties of liquid crystal polymers. To improve computational efficiency while preserving accuracy, we further simplify the full assembly of reactive mesogens to a dimer system with various conformations. These dimer conformations represent different degrees of $\pi$--$\pi$ stacking, and the dimer system energies adhere to the Boltzmann distribution. This approach is based on the Lebwohl–Lasher lattice model.~\citep{lebwohl1972nematic} This simple spin lattice model, rooted in classical physics, is particularly suitable for high-throughput screening due to its computational efficiency. The model utilizes the Monte Carlo method to simulate phase behavior by stochastically altering the orientations of individual molecules, with the Boltzmann distribution governing the process. Studies have shown that it effectively captures the first-order phase transition between ordered and disordered states.~\citep{lebwohl1972nematic} 

Initially, 50 reactive mesogen conformers were generated from the SMILES of input molecular structure via RDKit~\citep{weininger1988smiles, rdkit}, followed by structural optimization using semi-empirical GFN2-xTB method~\citep{bannwarth2019gfn2} with Crest.~\citep{pracht2020automated, pracht2024crest, grimme2019exploration} Molclus isostat tool~\citep{lu2016molclus} was used to merge identical conformers by applying a geometry threshold of 0.25 \AA\ and an energy threshold of 0.5 kcal/mol. Next, 200 dimers were generated from the selected conformers, with inter-dimer distances controlled to within 2 van der Waals radii, and random relative spatial orientations were assigned using Molclus genmer tool.~\citep{lu2016molclus} Dimers were further structurally optimized using the GFN2-xTB method in Crest, followed by the calculation of dimer probabilities at 400 K based on the Boltzmann distribution. Subsequently, the eight lowest energy dimers were identified among those with energy differences of less than 3 kcal/mol, serving as simplified computational units. This approach efficiently selects low energy conformers and captures the impact of intermolecular interactions on molecular geometry. Finally, quantum chemical calculations were performed to obtain the UV-Vis spectra and refractive index of the selected dimers, and the averaged properties were used to approximate the absorption and refractive index of the polymer film. The dimer generation workflow is illustrated in \autoref{fig:computational_workflow}, with the key steps highlighted in blue.

\subsection{UV-Vis Spectra Computation}
\subsubsection{Spectra Calculation and Experimental Validation}

TD-DFT has proven to be a reliable method for predicting the optical properties of organic molecules.~\citep{casida2012progress, adamo2013calculations, chen2017lateral} By applying traditional DFT to time-dependent systems, TD-DFT allows for accurate modeling of excited states, which are crucial for understanding electronic transitions in organic materials.~\citep{adamo2013calculations} To evaluate the absorption of liquid crystal polymers using the dimer pipeline, UV-Vis spectra for five commercial reactive mesogens, HCM-008 (RM257), HCM-009 (RM82), HCM-020 (RM23), HCM-021 (RM006/RM105), and HCM-083, were calculated with TD-DFT at the PBE0/6-31G(d,p) level. Concurrently, experimental measurements were conducted using UV-Vis spectroscopy. The results show that the computed values deviate by an average of only 7 nm from the experimental data, indicating a high degree of consistency with experimental observations. All calculations were performed using the Gaussian 16.c software package,~\citep{g16} and the simulated spectra are presented alongside the experimental data in~\autoref{fig:spectra-all}.

\begin{figure*}[!htb]
     \centering
     \includegraphics[width=\textwidth]{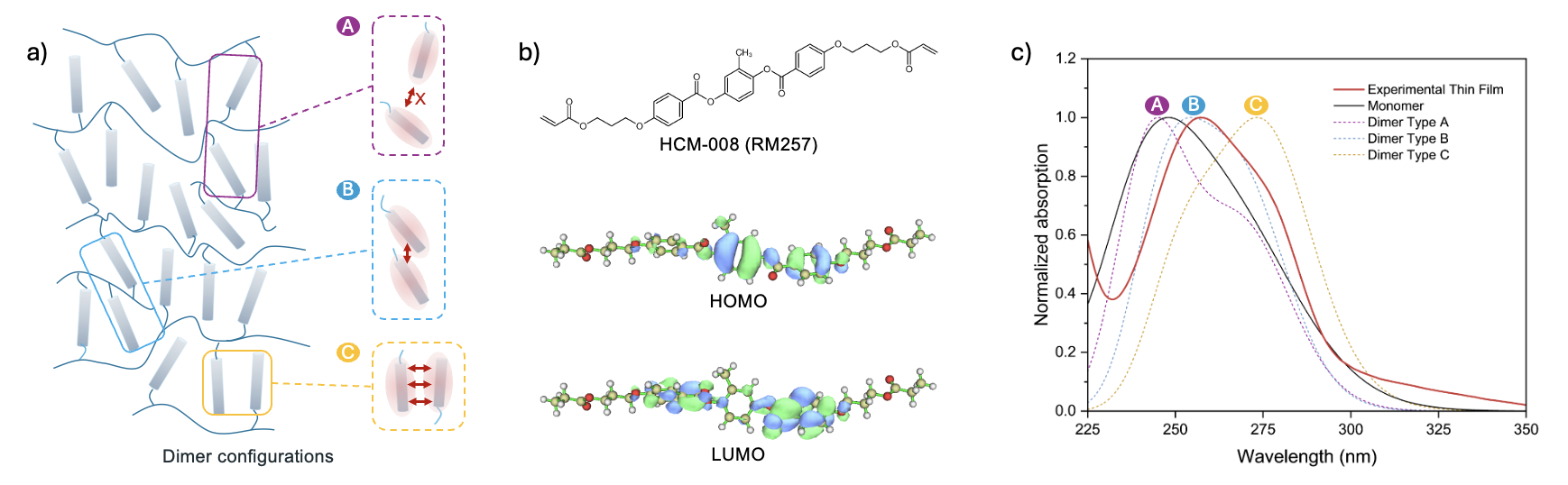}
     \caption{(a) Schematic illustration of a liquid crystal polymer network featuring three distinct dimer conformations, each representing varying degrees of $\pi$--$\pi$ stacking interactions. (b) Structure of HCM-008 and its HOMO-LUMO orbitals. (c) UV absorption spectra of HCM-008 from experimental monomer solution, polymer thin film and computational results of different dimer types.}
    \label{fig:uv_vis_hcm008}
\end{figure*}

As shown in the schematic drawing in \autoref{fig:uv_vis_hcm008}(a), the conjugated mesogenic cores are distributed in a liquid crystal polymer network connected by non-conjugated spacers. This arrangement results in dimer conformations with varying degrees of $\pi$--$\pi$ stacking. We classify these into three types: Type A represents dimers with almost no $\pi$--$\pi$ stacking interactions, Type B represents dimers with weak $\pi$--$\pi$ stacking interactions, and Type C represents dimers with the strongest $\pi$--$\pi$ interactions. As observed in both computational and experimental data, different degrees of $\pi$--$\pi$ stacking are critical in influencing optical responses, often leading to pronounced red-shifted absorption spectra or notable spectral broadening. To discuss in more detail, we take HCM-008 as an example, as shown in \autoref{fig:uv_vis_hcm008}(b). In HCM-008, the calculated orbital contributions indicate that the dominant allowed transition occurs at 250 nm, corresponding to a HOMO-LUMO transition identified as a $\pi \to \pi^*$ excitation within the molecule. \autoref{fig:uv_vis_hcm008}(c) illustrates that Type A, characterized by the weakest $\pi$--$\pi$ interactions, exhibits spectra closely resembling those of isolated monomers. In contrast, Type C, with the strongest $\pi$--$\pi$ interactions, displays the most significant red shift in its absorption spectrum. The combined effects of different dimer interactions are a primary factor contributing to the broadening of absorption peaks in polymer films. These findings highlight the importance of dimer conformations in determining the optical properties of liquid crystal polymers and support our approach of simplifying the simulation of UV-Vis spectra in liquid crystal polymers through the dimer configuration of the mesogenic core. A feasible approach to obtain a UV-Vis spectrum that closely approximates the optical transparency of the polymer thin film is to average the spectra of multiple dimer conformations.

\begin{figure*}[htbp]
     \centering
     \includegraphics[width=\textwidth]{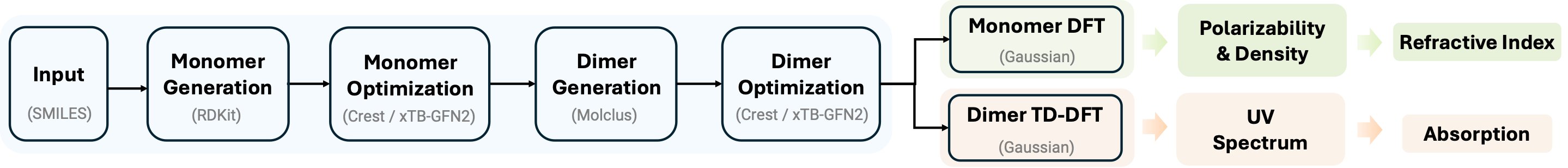}
     \caption{Computational pipeline for dimer generation and optical property calculations.}
     \label{fig:computational_workflow}
\end{figure*}

Building on the averaged spectra strategy, we developed an automated computational pipeline (\autoref{fig:computational_workflow}) to perform TD-DFT calculations on the top-ranked conformer candidates identified earlier, generating an averaged spectrum. This framework enables efficient, high-throughput molecular absorption calculations, facilitating subsequent genetic algorithm-based optimization. Beyond streamlining the computational process, this approach provides deeper insights into how mesogenic core interactions in the nematic phase influence the optical properties of liquid crystal polymers.

\subsubsection{Validation via Molecular Dynamics}

To further validate the accuracy of our computational pipeline based on dimer simplification, we conducted additional MD simulations, which are known for their ability to capture complex molecular interactions and phase transitions.~\citep{allen2019molecular, wilson2005progress, mcbride1999molecular} The phase behavior of five reactive mesogens, HCM-008 (RM257), HCM-009 (RM82), HCM-020 (RM23), HCM-021 (RM006/RM105), and HCM-083, was captured using MD simulations using Schrödinger’s Materials Science Suite with the OPLS4 force field (details in \autoref{appendix:MD}).~\citep{lu2021opls4, schrodinger2024, schrodinger2024force} In this process, all molecules successfully reproduced the nematic liquid crystal phase, validating the robustness of the protocol (\autoref{fig:Benchmark-p2}). 

We extracted dimer conformations by randomly selecting 200 dimers from the initialized liquid crystal phase (details in \autoref{appendix:MD}), ensuring an intermolecular distance of less than 4 \AA\ based on the nearest-neighbor criterion. Single-point energy calculations were then performed on these dimers using GFN2-xTB in Crest. The energy distribution of the dimers was found to approximate a Gaussian distribution, providing additional evidence of the distinct intermolecular interactions characteristic of the liquid crystal phase (\autoref{fig:distribution}). 


The average UV-Vis spectra were compared with experimental data (\autoref{tab:uv-table}). The results showed a strong correlation between the computed and experimental values, with an average deviation of only 6.2 nm. Similarly, TD-DFT calculations on the eight dimers generated via the designed dimer pipeline also yielded average UV-Vis spectra that closely matched the experimental results, with the same average error of 9.4 nm. 

Compared with the maximum absorption wavelength obtained from the MD-extracted dimer conformations, our designed dimer pipeline exhibits a red-shifted maximum absorption wavelength. This difference arises because the pipeline selects dimers based on their lowest energy conformations, resulting in stronger $\pi$--$\pi$ stacking and lower energy than MD-generated dimers, which follow a Gaussian distribution. Although this discrepancy arises from the pipeline's inability to fully capture the MD-based Gaussian distribution, the error is acceptable given the substantial acceleration of spectral calculations achieved through dimer simplification. These dimers will also serve as a basis for further quantum chemical calculations of refractive indices, providing valuable support for subsequent material screening efforts.

\subsection{Refractive Index Computation}

The macroscopic refractive indices of liquid crystals are typically derived from the Lorentz-Lorenz equation. Due to the anisotropic nature of extraordinary ($n_{e}$) and ordinary ($n_{o}$) refractive indices, Vuks \citep{vuks1966optical} introduced a semi-empirical equation that incorporates the isotropic local field, relating microscopic molecular polarizability to the macroscopic refractive indices in anisotropic media.~\citep{zakerhamidi2012order, alipanah2022temperature}
\begin{equation}
\frac{(n_{e,o}^2 - 1)}{\langle n^2 \rangle + 2} = \left( \frac{4\pi}{3} \right) N \alpha_{e,o}\ ,
\end{equation}
where $N$ is the number of molecules per unit volume, $\alpha_{e}$ and $\alpha_{o}$ are the respective molecular polarizabilities, and ${\langle n^2 \rangle}$ is defined as:
\begin{equation}
    \langle n^2 \rangle = \frac{n_e^2 + 2n_o^2}{3}\ .
\end{equation}
The ratio of normalized polarizability can be expressed as:
\begin{equation}
    \frac{\alpha_e}{\alpha_o} = \frac{n_e^2 - 1}{n_o^2 - 1}\ .
\end{equation}

Therefore, by calculating the molecular anisotropic polarizability and number density, the refractive index can be accurately predicted. The following sections provide a detailed explanation of how these results are calculated through first-principles methods, complemented by a calibration against experimental data.

\subsubsection{Number Density}

The number density is typically determined from the density of MD simulation trajectory. However, this approach is both computationally expensive and time-consuming, posing challenges for high-throughput material screening. To address this, current methods that substitute molecular volume for MD-calculated density have demonstrated strong correlations with experimental results for amorphous polymers and liquid systems, offering a more cost-effective alternative to traditional MD simulations.~\citep{afzal2018combining}

To validate that the molecular volume can be applied to the density calculations of liquid crystals, we investigated the temperature-dependent density behavior of the five reactive mesogens. The results showed that while density decreased with temperature, the slope of density change remained nearly constant (\autoref{fig:Density-Temperature}). This consistency arises from the highly uniform ellipsoidal shapes of reactive mesogen molecules, which, despite variations in molecular weight, exhibit similar packing coefficients in the nematic phase. This finding supports the feasibility of using molecular volume to accelerate number density calculations.

We employed Bader's definition of molecular volume in the condensed phase, where molecular volume is defined using the electron density isosurface as the van der Waals surface, accounting for volume deformation due to electronic effects.~\citep{bader1987properties, liu2021intermolecular} The molecular volume was computed using the Marching Tetrahedra method on grid data, as implemented in Multiwfn software.~\citep{lu2012quantitative, lu2024comprehensive} A comparison of the densities from the molecular volumes of monomers and dimers for the five reactive mesogens revealed a less than 2\% difference in the densities obtained from wavefunctions. (\autoref{tab:dimer-RI-table}). Thus, using the molecular volume of monomers significantly enhances computational efficiency, while maintaining desired accuracy.

Subsequently, we compared the nematic phase densities at 298 K of 77 literature-reported liquid crystal molecules obtained via MD simulations with densities derived from molecular volumes calculated using DFT wavefunctions (\autoref{fig:density-benchmark}). The results showed a strong correlation ($R^2=0.92$) between the two methods. This indicates that using the molecular volume of reactive mesogens can greatly simplify and accelerate the number density calculations.

\subsubsection{Polarizability Computation}

Polarizability ($\alpha$) of liquid crystal molecules can be derived from wave function-based methods and DFT.~\citep{helgaker2012recent, hait2018accurate} By using the linear response theory, polarizability is computed by analyzing the system’s reaction to an external perturbation, such as a constant electric field.~\citep{stott1980linear} The first-order response of the zero-field energy $E(0)$ is the molecular dipole moment, reflecting the electron density distribution in polar molecules. The second-order response is the static polarizability tensor ($\alpha$), which quantifies the deformation of the electron cloud under the external field. Due to the rod-like structure and $\pi$-conjugated backbone in liquid crystal molecules, the polarization along the molecular long axis is typically greater than that along the short axis. This structural anisotropy results in a directionally dependent dipole response to external electric fields. Therefore, the extraordinary polarizability $\alpha_{e}$  and ordinary polarizability $\alpha_{o}$ are defined as:
\begin{equation}
    \alpha_{e} = \alpha_{xx}\ ,
\end{equation}
\begin{equation}
    \alpha_{o} = \frac{\alpha_{yy} + \alpha_{zz}}{2}\ ,
\end{equation}
where \(\alpha_{xx},\ \alpha_{yy},\ \alpha_{zz}\) represent the polarizability components in the \(x\), \(y\), and \(z\) directions, respectively.

The polarizability is a frequency-dependent property. However, only the static polarizability was calculated to derive the refractive index under zero-field conditions. Given that the target liquid crystal molecules are transparent with negligible absorption in the visible region and lack low energy excited states, the frequency-dependent polarizability remains nearly constant or slightly monotonically decreases. It ultimately converges to the static polarizability value.~\citep{jansik2004calculations} Consequently, the relative refractive indices of different molecular structures remain unaffected by frequency, thus not impacting the subsequent material screening process. Moreover, the computational cost of determining static polarizability is significantly lower than that of frequency-dependent polarizability, enabling the use of static values to approximate the experimentally measured refractive indices at specific wavelengths.

Accurately calculating the polarizability is a computationally intensive task that imposes strict requirements on the dispersion functions of the basis set and the selection of DFT functionals.~\citep{baranowska2015applicability, hickey2014benchmarking, afzal2019benchmarking} To reduce computational expense and improve efficiency, the ZPol basis set has been rigorously validated as a reliable choice to achieve moderate accuracy in the evaluation of molecular linear electrical properties.~\citep{baranowska2007reduced, benkova2005reducedsize, benkova2005reduced} It has demonstrated competitiveness with larger and more general basis sets while maintaining reliability in polarizability estimates for organic molecules.~\citep{lee2012density, baranowska2012accurate} To further enhance computational efficiency, we calculated the polarizability of dimers for the five reactive mesogens and their constituent monomers separately, with results summarized in~\autoref{tab:dimer-RI-table}. The findings show that the averaged polarizability of the two monomers deviates by less than 2\% from that of the corresponding dimer, demonstrating that monomers can serve as a computationally efficient alternative for polarizability calculations. Furthermore, conformational differences between dimers and monomers contribute to an error of less than 2\%. Therefore, we select the lowest energy dimer conformation, compute the polarizability of its individual reactive mesogens separately, and use the averaged value.

We used the M06-HF functional~\citep{zhao2006density} along with the ZPOL basis set (containing both polarization and diffuse functions) to calculate the polarizabilities of 77 liquid crystal small molecules reported in the literature, and compared the results with experimental values, as shown in~\autoref{fig:RI-all}. All calculations were performed using the Gaussian 16.c software package.~\citep{g16} The $n_{avg}$ calculated by $\alpha_{avg}$ showed excellent agreement with the experimental results, while the calculated $n_{e}$ was slightly higher and the $n_{o}$ was slightly lower than the experimental values. This discrepancy arises because DFT calculations for individual molecules do not account for the influence of order parameter on $n_{e}$ and $n_{o}$ in liquid crystal polymer films. Consequently, in the subsequent genetic algorithm-based structure exploration, we used $n_{avg}$ as the criterion to identify candidates with high refractive index. 

In the high-throughput refractive index calculation pipeline, DFT calculations are performed on monomers derived from the lowest energy dimer configuration. Molecular volume is obtained from wavefunctions, and density is simulated using a calibrated packing coefficient. By integrating these parameters with polarizability calculations, the polymer film's refractive index is approximated. This approach optimally balances computational efficiency and accuracy, making it particularly suitable for high-throughput screening and the systematic exploration of liquid crystal materials.

\section{Genetic Algorithm Design}

A major challenge in designing novel materials for optical applications is identifying optimal molecular structures within the vast chemical space. To overcome this challenge, we use an inverse design strategy based on a genetic algorithm for global multi-objective optimization - a widely recognized approach for accelerating the discovery of organic materials with desired optical and electronic properties.~\citep{suh2020evolving, hiener2022pareto, abarbanel2023using} Compared to machine learning, genetic algorithms are computationally efficient, as they do not require large training datasets, and they offer a significant speed advantage over traditional experimental workflows.  During multiple iterations of the evolutionary algorithm, the population of monomer structures undergoes computational crossover and mutation operations (mimicking biological processes), such as the exchange or substitution of molecular fragments. Ultimately, candidates that pass the initial screening are subjected to further property computations with first-principles methods in our computational pipeline.

\begin{figure*}[!htb]
     \centering
     \includegraphics[width=\textwidth]{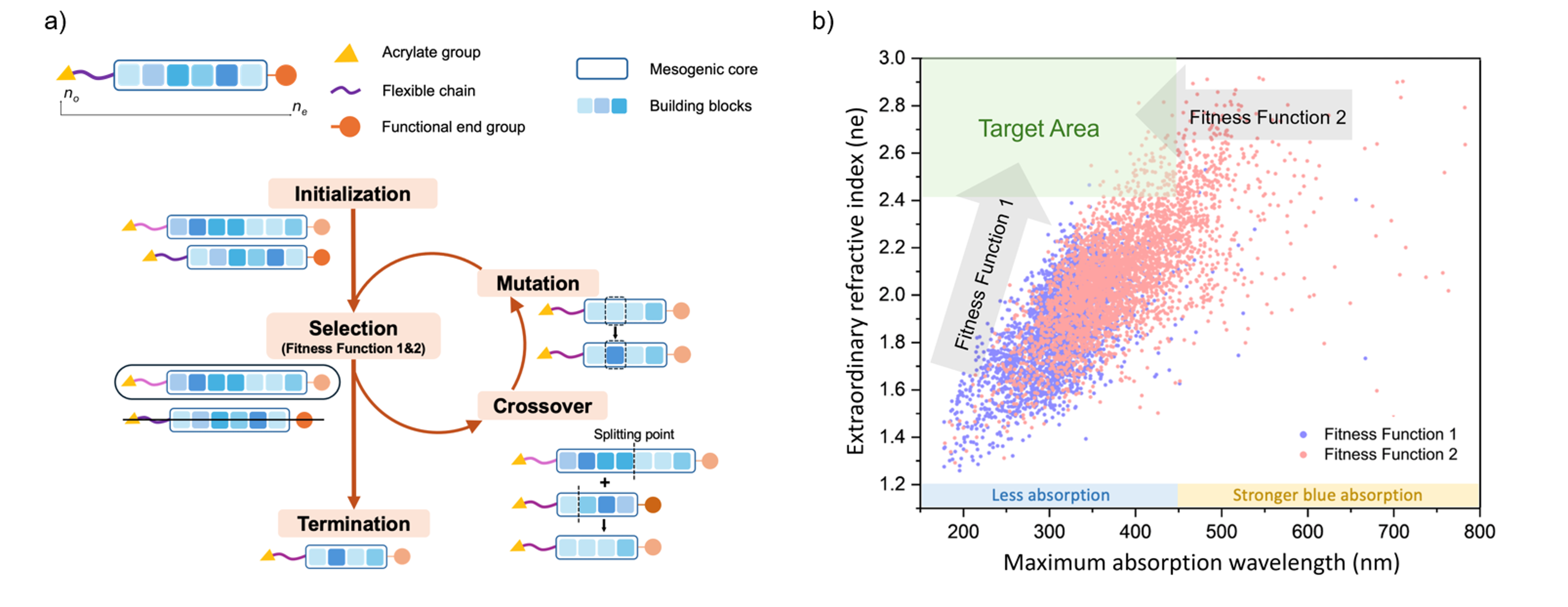}
     \caption{(a) Design of reactive mesogen structures and the five steps of genetic algorithm for evolutionary iterations: initialization, selection, crossover, mutation, and termination. (b) Distribution of maximum absorption wavelength and $n_{e}$ of molecules obtained from Fitness functions 1 and 2.}
     \label{fig:GA-all}
\end{figure*}

The initial batch of generated reactive mesogen structures begins with the iterative process of the genetic algorithm. Using an established computational pipeline, UV-Vis spectra and theoretical refractive index values are obtained for each molecule. A fitness function ranks the molecules, selecting those that best balance transparency and refractive index for the next iteration. New molecular structures are generated through crossover and mutation, and the process continues until the generated molecules meet the optical requirements. The iteration then terminates, yielding the desired reactive mesogen structures (\autoref{fig:GA-all}(a)).

The highlight of our pipeline is the molecular generation from building blocks, which is unique to liquid crystals, which will be discussed in more detail in the following sections.

\subsection{Molecular Building Blocks}

The molecular building blocks include the mesogen core, an acrylate-linked alkyl chain, and a terminal substituent, as illustrated in~\autoref{fig:GA-all}(a). Each moiety types was identified and extracted from a liquid crystal molecule literature.

The database comprises four polymerizable alkyl chain building blocks incorporating different heteroatoms, 15 substituent building blocks positioned at the opposite end of the mesogen, and 52 mesogen core building blocks, including 16 linear bridges and 36 aromatic rings (\autoref{fig:BB-database}). The length of the mesogen core is constrained to 3-7 units, as shorter cores exhibit insufficient anisotropy between their long and short axes, hindering liquid crystal phase formation. However, excessively long cores exhibit extended conjugation, which significantly reduces optical transparency. The unique design of these building blocks for reactive mesogens precisely captures key molecular structural features, enhancing the efficiency of the genetic algorithm by accelerating convergence toward target properties. The liquid crystalline phase behavior of the generated molecules is further validated through MD simulations, as discussed in subsequent sections.

\subsection{Fitness Functions}

According to the sum-over-states (SOS) equations, achieving both high optical transparency and high polarizability in a single monomer is inherently limited by electronic delocalization. Molecules with higher transparency per unit volume typically exhibit lower polarizability, as the delocalized electrons are more easily excited by shorter wavelengths. Given these inherent trade-offs, where low absorption and high refractive index often conflict, optimizing these characteristics simultaneously becomes a complex challenge. Therefore, the fitness function plays a critical role in ranking, selecting, and filtering target structures, ensuring that the selected molecules are aligned with the desired target properties.

To optimize both optical transparency and refractive index, we designed two specialized fitness functions to guide the iterative selection process. The fitness functions consist of three components: the transparency score, the average refractive index value score, and the anisotropic refractive index difference score. By adjusting the weights of each component, we refine the ranking of candidates at each iteration, ensuring that only the most promising candidates progress to the next stage.

The transparency score is determined by the maximum absorption wavelength and the ratio of the absorbance of a molecule at its maximum absorption wavelength to its absorbance at 460 nm. Molecules with shorter maximum absorption wavelengths and lower absorbance at 460 nm receive higher transparency scores. The average refractive index score prioritizes molecules with higher average refractive indices, promoting the selection of high refractive index candidates. The anisotropic refractive index difference is included to avoid deviations from a rod-shaped molecular shape. Randomly generated reactive mesogen core building blocks may contain multiple bridge linkers, which might reduce molecular rigidity. This can hinder nematic phase packing or even prevent the molecule from being considered a liquid crystal monomer. Including the anisotropic refractive index difference in the fitness function helps favor rod-like molecules, making it more likely that they will go to the next iteration.

Fitness function 1 gives more weight to the transparency score. It starts with transparent molecules, especially those with a maximum absorption wavelength shorter than 460 nm, selecting candidates with higher refractive indices in subsequent iterations. Fitness function 2 focuses on molecules with refractive indices greater than 2.0, filtering for transparency in later stages. The specific formulas for both fitness functions can be found in \autoref{appendix:fitness function}. Although the two functions have different weightings, both prioritize molecules that achieve high refractive index and transparency, making it more likely for them to progress to the next iteration. As shown in \autoref{fig:GA-all}(b), molecules optimized with Fitness function 1 tend to cluster in regions with low absorption and low $n_{e}$, while those optimized with Fitness function 2 are found in areas with higher $n_{e}$ and closer to the visible absorption range. This highlights the challenge of balancing transparency and refractive index, and demonstrates how our dual fitness functions effectively guide molecules toward the desired optical properties. With each iteration, the molecules move closer to our target area.

\section{Results and Discussion}

\subsection{Genetic Algorithm Convergence}

\begin{figure*}[htbp]
     \centering
     \includegraphics[width=\textwidth]{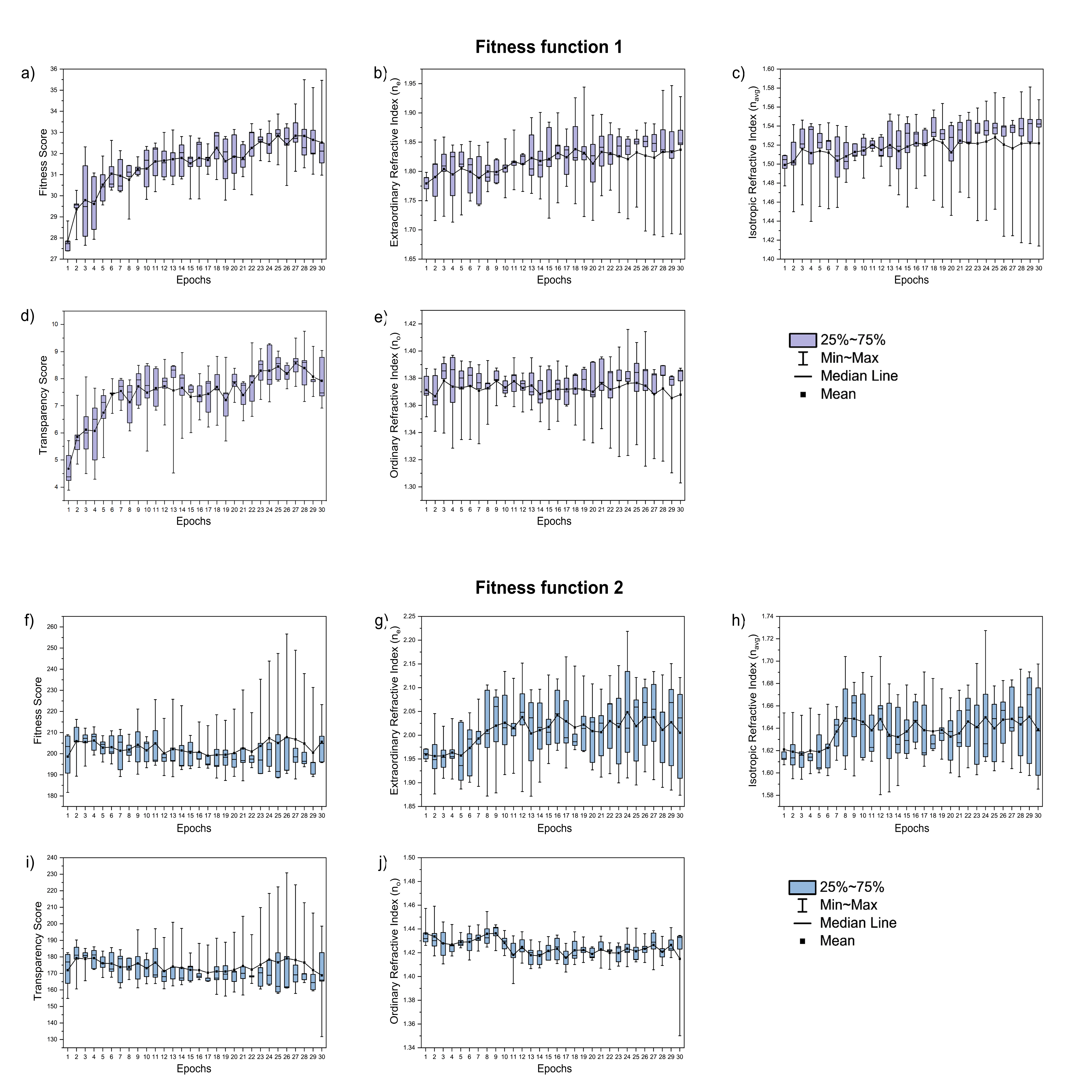}
     \caption{Score and property distributions across all five independent genetic algorithm runs: (a) and (f) show fitness score, (b) and (g) show extraordinary refractive index $n_{e}$, (c) and (h) show average refractive index $n_{avg}$, (d) and (i) show transparency score , (e) and (j) show ordinary refractive index $n_{o}$ for Fitness functions 1 and 2, respectively.}
     \label{fig:fitness-scores-all}
\end{figure*}

\begin{figure*}[htbp]
     \centering
     \includegraphics[width=\textwidth]{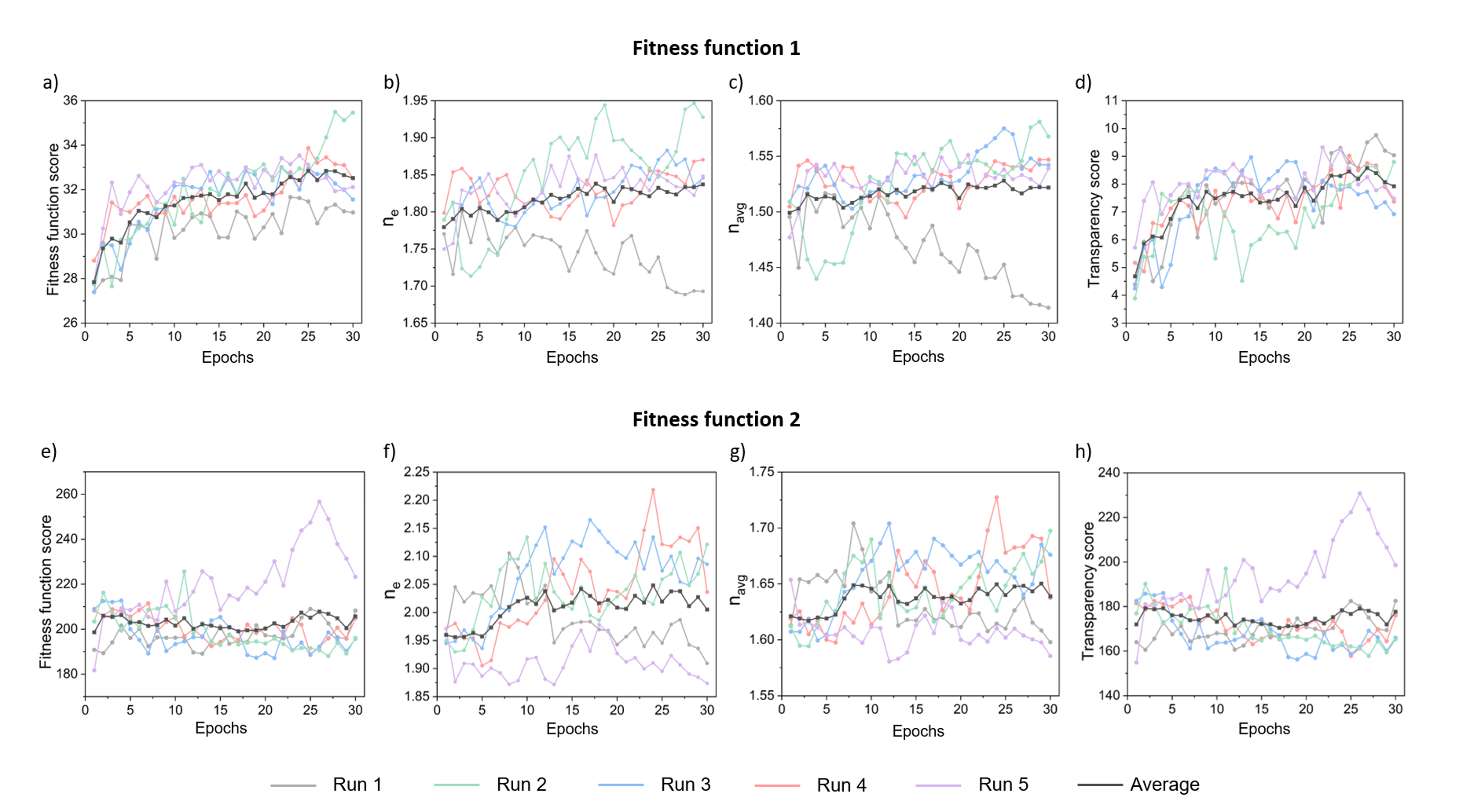}
     \caption{Metrics across each of the five independent genetic algorithm runs: (a) and (e) show fitness score, (b) and (f) show extraordinary refractive index $n_{e}$, (c) and (g) show average refractive index $n_{avg}$, (d) and (h) show transparency score for Fitness functions 1 and 2, respectively.}
     \label{fig:convergency}
\end{figure*}

For both fitness functions, we performed five independent runs of the genetic algorithm, starting from a random initial seed of 32 molecules, each run lasting 30 epochs. To evaluate the convergence behavior of the genetic algorithm, we analyzed the evolution of fitness scores, refractive indices, and transparency scores over epochs for each run. We report the individual and combined behaviors of the runs shown in \autoref{fig:fitness-scores-all} and~\autoref{fig:convergency}.

Fitness function 1 showed minimal variability across the five runs, and all runs demonstrated a clear convergence toward maximizing the fitness score. The maximum, minimum, and average scores of the population consistently increased, indicating that the genetic algorithm successfully identified molecules with both high transparency and refractive index (\autoref{fig:fitness-scores-all}(a)). Transparency scores within the population also showed significant convergence toward the maximum value of 10, with high consistency across runs (\autoref{fig:fitness-scores-all}(d)). Moreover, both $n_{e}$ and $n_{avg}$ showed an increase across epochs in all five runs, with the average values improving from 1.80 to 1.95 for $n_{e}$ and from 1.51 to 1.58 for $n_{avg}$ (\autoref{fig:fitness-scores-all}(b) and \autoref{fig:fitness-scores-all}(c)). In contrast, $n_{o}$, representing the refractive index along the short axis of the reactive mesogen, remained unchanged during the molecular design process (\autoref{fig:fitness-scores-all}(e)). Given the competing relationship between transparency and refractive index, Fitness function 1 prioritizes maximizing molecular transparency. As a result, the improvement in transparency scores during the iterations far outpaced the increases in $n_{e}$ and $n_{avg}$, which aligns with the expected outcome for the target molecules.

Fitness function 2 showed greater variability in the five independent runs compared to Fitness function 1. Both $n_{e}$ and $n_{avg}$ demonstrated increases during the initial ten epochs before convergence, with $n_{e}$ increasing from 1.95 to 2.03 and $n_{avg}$ increasing from 1.62 to 1.65 (\autoref{fig:fitness-scores-all}(g) and \autoref{fig:fitness-scores-all}(h)). Compared to Fitness function 1, Fitness function 2 achieved higher values for both $n_{e}$ and $n_{avg}$. Obvious variability was observed in the fitness function score, with the mean value remaining stable at approximately 205 throughout iterations, while the maximum score in one run reached as high as 260 (\autoref{fig:convergency}(e)). Similar variability was observed for the transparency score, where the average value remained nearly constant, but the maximum value improved significantly between epochs. These distinct variations in transparency scores can be attributed to the constraint imposed by Fitness function 2, requiring $n_{avg}$ to exceed 1.6. This restriction effectively narrows the search space, making advancements in transparency highly reliant on the stochastic discovery of top-performing candidates. As a result, the variability between different runs became more pronounced, with the success of the process relied on the inherent randomness of crossover and mutation in the genetic algorithm. Moreover, the design of the molecule along the short axis remained unchanged, so the $n_{o}$ value showed little fluctuation (\autoref{fig:fitness-scores-all}(j)). Despite the considerable variability in the optimization outcomes across different runs with Fitness function 2, the average refractive index between 5 runs consistently reached values above 2.0 for all runs. Furthermore, the refractive index demonstrated an upward trend with fluctuations in all three runs, aligning with our expectations.

\subsection{Building Block Frequencies}

\begin{figure*}[htb]
     \centering
     \includegraphics[width=\textwidth]{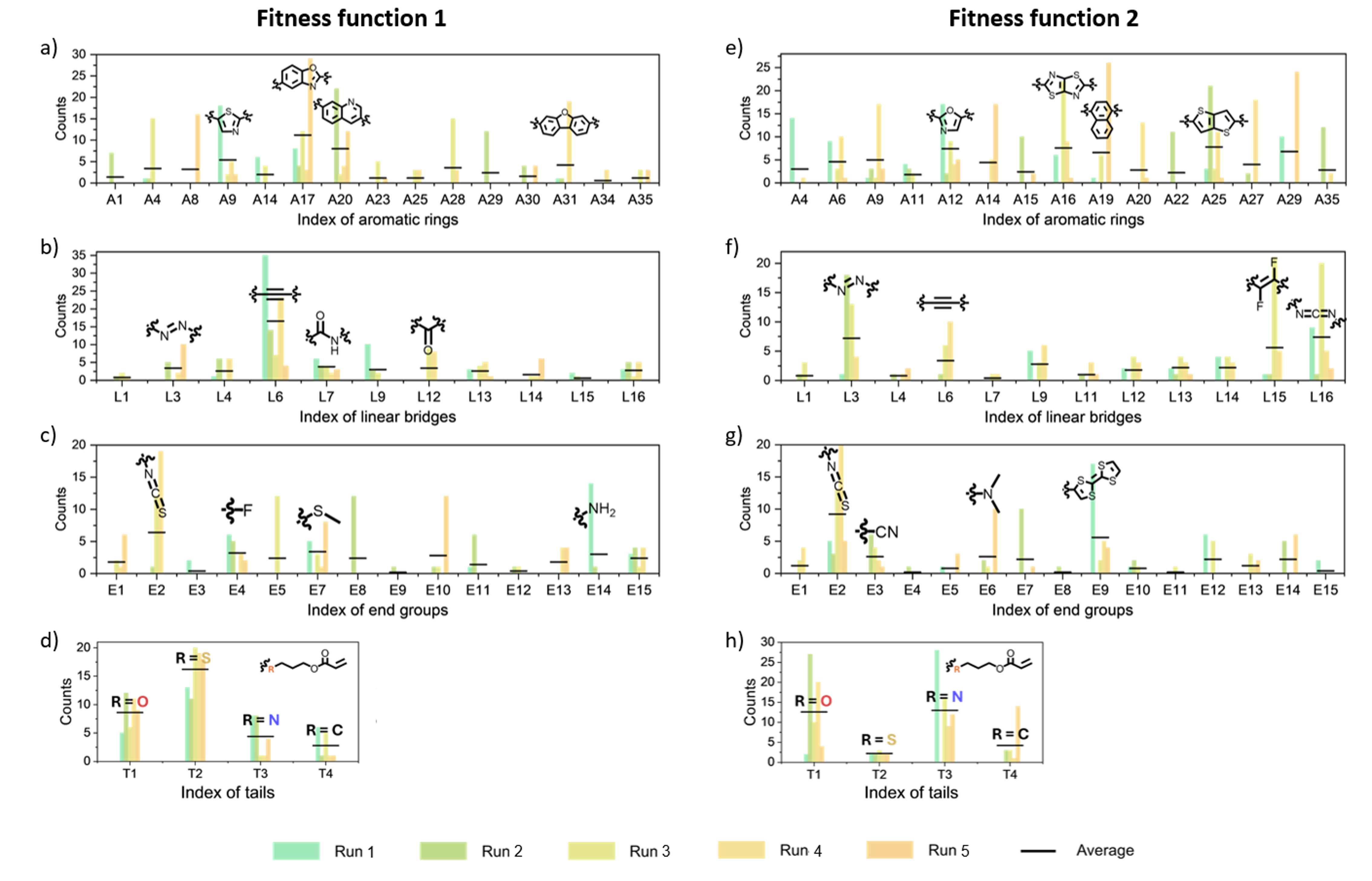}
     \caption{Building block frequencies of each of five genetic algorithm runs with Fitness functions 1 and 2, including (a) and (e) aromatic rings, (b) and (f) linear bridges, (c) and (g) end groups, (d) and (h) tails.}
     \label{fig:building-block-freq}
\end{figure*}

We analyzed the high-frequency building blocks in reactive mesogens optimized by the two fitness functions. In Fitness function 1, A17 (benzoxazole) and A20 (quinoline) consistently exhibited high occurrence in all five runs, whereas other aromatic building blocks showed high frequencies in only one run (\autoref{fig:building-block-freq}(a)). In Fitness function 2, A12, A16, and A25 were dominant in at least one run, with frequencies exceeding 50\% and remaining prevalent in others (\autoref{fig:building-block-freq}(e)). Despite the differing selection criteria of the two fitness functions, both favored aromatic rings with multiple heteroatoms and fused ring systems. The presence of heteroatoms modifies the electron distribution, influencing dipole interactions and electronic transitions, which can enhance polarizability. Similarly, fused rings with extended $\pi$-conjugation facilitate electron delocalization, further improving polarizability. However, excessive ring fusion or the introduction of strongly electron-withdrawing heteroatoms can shift absorption into the visible range, reducing optical transparency. These high-frequency building blocks provide a balance between polarizability and transparency.

For the bridge linkers in the mesogen core, C$\equiv$C triple bonds were highly frequent and consistent in all five runs in both fitness functions, especially in Fitness function 1, which prioritizes transparency. Their rigid structure stabilizes molecular geometry, creating a more structured backbone that enhances the packing efficiency of reactive mesogens. This improved packing promotes intermolecular interactions, contributing to greater order in the nematic phase. Additionally, the high bond energy and localization of $\pi$-electrons in triple bonds widen the HOMO-LUMO gap, shifting absorption into the UV region and avoiding visible absorption. As a result, C$\equiv$C triple bonds improve both transparency and structural rigidity, making them a preferred choice over other linear linkers. In contrast, Fitness function 2 favored azo (-N=N-) linkages and halogen-containing bridge linkers. Halogen substitution modifies electron density distribution and can enhance polarizability via induced dipole effects. However, azo bonds are not strictly linear and can undergo cis-trans isomerization under light exposure, potentially disrupting the nematic phase stability. Consequently, they were less effective in Fitness function 1, which prioritizes optical transparency. This highlights C$\equiv$C triple bonds as an optimal choice, offering a balance between high polarizability, rigidity, and excellent optical transparency.

The \texttt{-N=C=S} end-group, prevalent across all optimization runs, significantly improves both polarizability and transparency. The electron-withdrawing sulfur and nitrogen atoms, by reducing electron density, promote electron delocalization within the conjugated system, thereby enhancing the polarizability. Furthermore, conjugation between the \texttt{-C=N} and \texttt{-C=S} groups further facilitates electron delocalization, modifying the molecular electronic structure. This change in electronic structure shifts the absorption to the UV region, as the modified energy levels make electronic transitions more likely to occur in the UV range. Consequently, the \texttt{-N=C=S} end-group not only enhances polarizability but also maintains transparency, making it an ideal choice for the design of liquid crystal molecules.

For flexible chains with different heteroatoms, the results across all five runs were consistent in both Fitness functions 1 and 2. Molecules substituted with oxygen and nitrogen contributed more to polarizability, while sulfur played a more significant role in enhancing transparency. Sulfur, with its larger size and lower electronegativity, participating in conjugation through electron delocalization. This electron donation helps distribute electron density more evenly across the molecule, avoiding electronic transitions in the visible range. As a result, sulfur-containing flexible chains are more transparent, as their electronic transitions tend to shift into the UV range. In contrast, carbon atoms in the flexible side chain had the weakest impact on both polarizability and transparency across both fitness functions, likely due to their lower electron-donating ability and smaller size, which make them less effective at stabilizing the structure or shifting absorption toward shorter wavelengths.

These outcomes may be attributed to the inherent stochasticity of genetic algorithms, which require a greater number of iterations and repeated experiments to avoid local minima. The current limited iterations and repetitions may have overlooked top candidates. However, within the limited number of runs and epochs, we have still identified several consistent candidates, which have been highly valuable for the molecular design.

\subsection{Generated Material Validations}

To verify the existence of the nematic phase in the generated reactive mesogen structures, we selected ten candidates with $n_{e}$ values exceeding 2.0. Additionally, these candidates had absorption intensities below 0.1\% at 460 nm compared to the maximum absorption peak absorbance. The selected candidates were then analyzed using the MD cooling process (\autoref{fig:candidates-P2}). The results showed that six candidates exhibited a nematic phase within the temperature range of 200 K to 700 K. Candidate 1 did not undergo a nematic phase transition and maintained an order parameter below 0.2. This result can be attributed to the elongated aspect ratio of the mesogen core, which is entirely composed of C$\equiv$C triple bond bridge linkers (\autoref{fig:candidates-P2}). Compared to cores formed by aromatic rings, linear bridges produce a much smaller radius along the $n_{o}$ direction, hindering the formation of ordered structures during the cooling process. In contrast, Candidates 8-10 exhibited phase transition temperatures that exceeded 700 K. These candidates consistently incorporated the fused ring building block and the mesogen core composed of four building blocks, imparting greater rigidity and enhanced $\pi$--$\pi$ stacking interactions. Consequently, the increased molecular rigidity and improved stacking efficiency substantially elevated the transition temperatures of these candidates.

Among the six molecules exhibiting liquid crystal phases, we further selected Candidate 2, which had the lowest transition temperature, for polarizability density analysis to investigate the relationship between high refractive index structures and their properties. The results revealed that the polarizability density was primarily concentrated near the triple bonds and the thiocyanate group at one end of the molecule, as well as around the sulfur atom at the opposite end. This localization can be attributed to the high electron density and polarizability of the triple bonds and the strong dipole moment of the thiocyanate group, which enhances the responsiveness of the molecule to external electric fields. Similarly, the sulfur atom contributes significantly to the polarizability due to its large atomic radius and high polarizability, further emphasizing its role in influencing the optical properties of Candidate 2.

\begin{figure*}[!htbp]
     \centering
     \includegraphics[width=\textwidth]{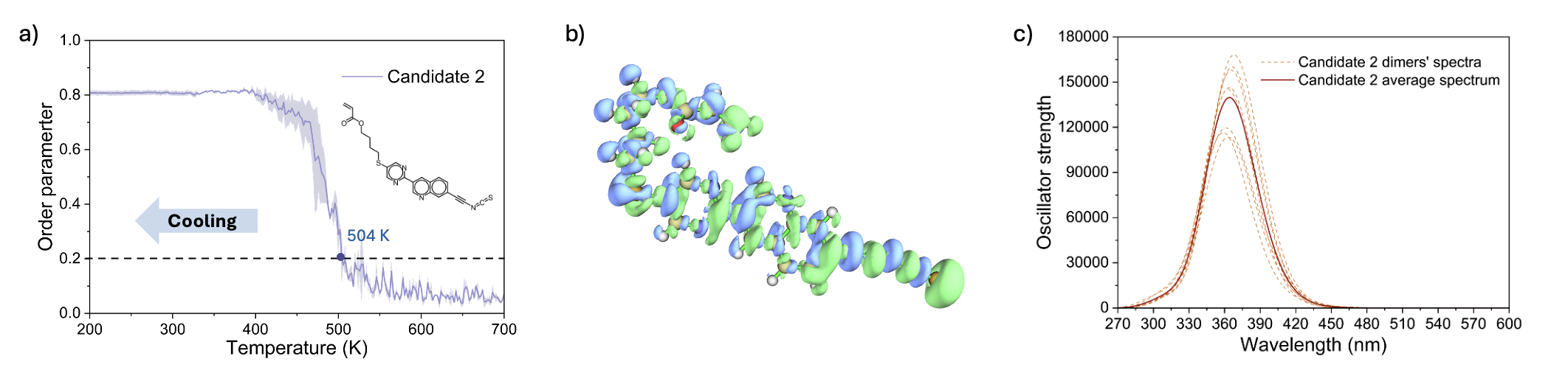}
     \caption{(a) Order parameter as a function of temperature for Candidate 2 during the MD cooling process. (b) Polarizability density distribution of Candidate 2, with the green and blue isosurfaces representing positive and negative values, respectively. (c) Average and individual spectra of eight dimers of Candidate 2 selected through the dimer generation pipeline.}
     \label{fig:candidate2-all}
\end{figure*}

Subsequently, we employed the TD-DFT method to calculate the excitation energies and oscillator strengths of the first 20 excited states of Candidate 2. The maximum absorption peak was observed at 365 nm (\autoref{fig:candidate2-all}). A strong absorption band is attributed to the doubly degenerate HOMO-1 $\rightarrow$ LUMO and HOMO $\rightarrow$ LUMO+1, which correspond to the $\pi \to \pi^*$ transition, while all other excitations contribute negligibly to the absorption spectrum (\autoref{fig:candidate2-orbitals}). As no significant absorption band is observed in the visible region, the polymer formed by Candidate 2 is expected to be colorless.

\section{Conclusion}

In this study, we developed a material discovery pipeline for liquid crystal molecules that integrates the semi-empirical GFN2-xTB method with first-principles (TD-)DFT calculations. This integration enables rapid iterations within a genetic algorithm framework, allowing for efficient exploration of the chemical search space. Our approach successfully identifies potential reactive mesogen structures characterized by high transparency and high refractive indices.

The dimer generation pipeline incorporates the influence of liquid crystal molecular packing. We have validated the accuracy and efficiency of this pipeline by comparing experimental and calculated values for five commercial reactive mesogens. These comparisons demonstrate the pipeline's capability to reliably predict material properties, thereby supporting its use in the discovery of new liquid crystal materials.

The genetic algorithm demonstrated its effectiveness as a methodology for optimizing material absorption and refractive index by efficiently exploring and targeting specific regions within the chemical search space. Despite the inherent randomness of genetic algorithms, it consistently identified recurring high-frequency building blocks and specific structural arrangements across independent runs. This consistency suggests that increasing the number of iterations and independent runs will further enhance the robustness and reliability of the results, solidifying the effectiveness of this design strategy.

Our computational analysis revealed that the 10 screened candidates exhibit less than 0.1\% absorption at 460 nm compared to the maximum absorption peak absorbance, with $n_{e}$ values exceeding 2.0. The highest candidate shows an $n_{e}$ value exceeding 2.1, and all candidates were further validated with MD simulations to confirm the existence of the nematic phase. In cases with lower transparency constraints or additional side-chain modifications, we anticipate even higher $n_{e}$ values for our candidates.

Future research could focus on improving the computational efficiency of the pipeline and advancing the prediction of nematic phase transitions. These advancements would enable deeper exploration of the chemical space for materials with enhanced optical properties. Such efforts would expand the potential applications of liquid crystal-based optical materials and further the development of high-performance optical devices.

\bibliographystyle{assets/plainnat}
\bibliography{ref-fair-srt}

\clearpage
\newpage
\beginappendix
\setcounter{figure}{0}
\renewcommand{\thefigure}{S\arabic{figure}}

\section{Molecular Dynamics Simulations}
\label{appendix:MD}
\begin{figure*}[htbp]
     \centering
     \includegraphics[width=0.9\textwidth]{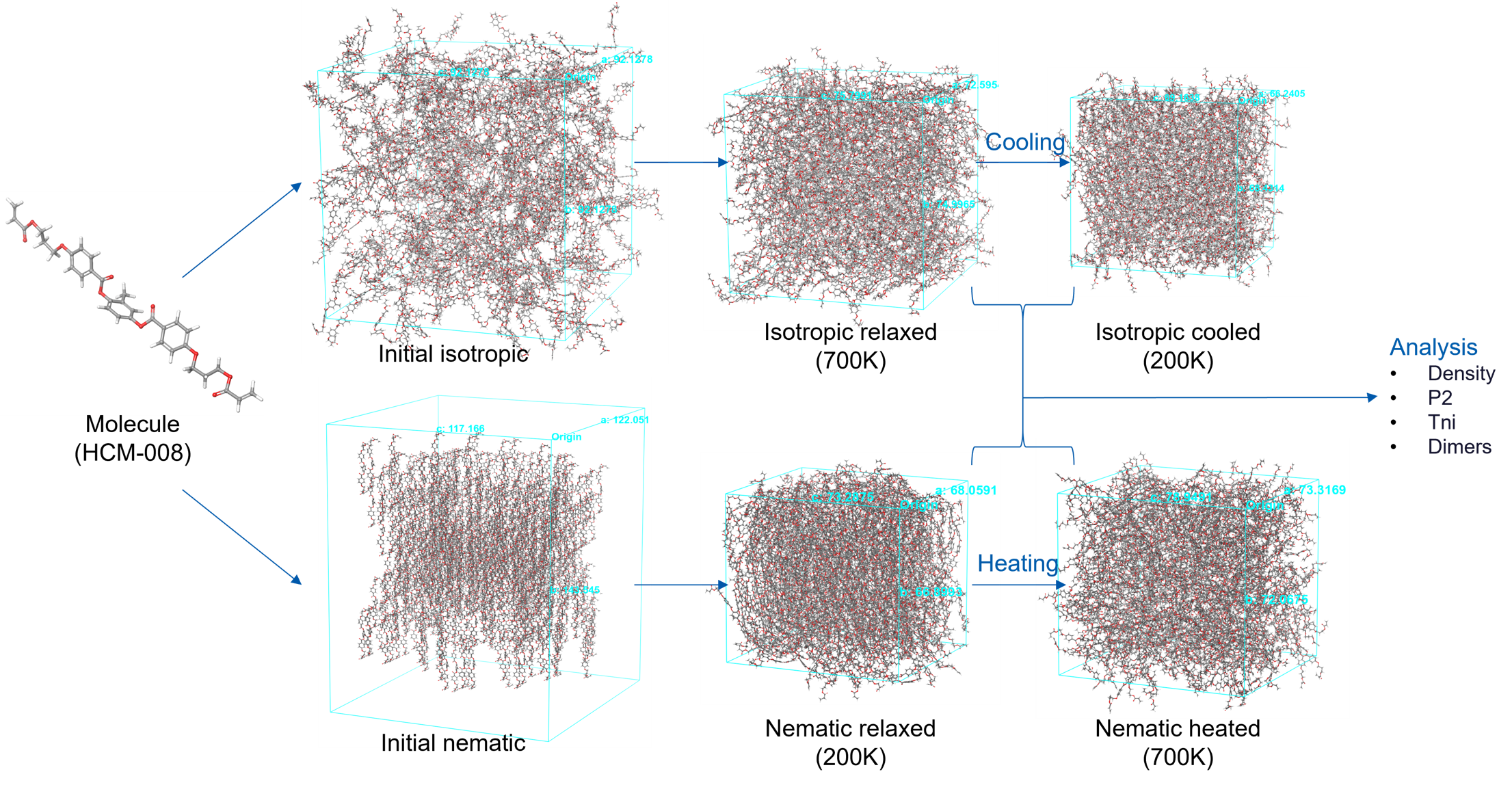}
     \caption{Molecular dynamics cooling and heating protocol for the nematic phase transition computations.}
     \label{fig:MD-workflow}
\end{figure*}

All computational systems for molecular dynamics (MD) were built and simulated using Schrödinger’s Materials Science Suite Release 2024-3, employing the OPLS4 force field.~\citep{lu2021opls4, schrodinger2024, schrodinger2024force} Initially, molecules of a liquid crystal candidate were packed into a simulation cell using the disordered system builder to create a disordered cell, with approximately 30,000 atoms and a density of 0.5 g/cm$^3$. To ensure statistical reliability, two sets of randomly packed simulation cells were generated for each candidate, with simulations run on both sets.

For equilibration, MD simulations were performed at 700 K. The procedure included 20 ps of Brownian dynamics in NVT ensemble (number of atoms [N], volume [V], and temperature [T] are held constant) at 10 K, followed by 20 ps of Brownian dynamics in NPT ensemble (number of atoms, pressure [P], and temperature are held constant) at 1 atm and 100 K, then 100 ps MD simulation in NPT ensemble at 1 atm and 300 K, and finally 10 ns MD simulation in NPT ensemble at 1 atm and 700 K. It was ensured that the potential energy and density of systems had reached a consistent average value at the end of the equilibration procedure. We used the Nosé–Hoover thermostat and the Martyna–Tobias–Klein barostat during this process.~\citep{nose1984unified, hoover1985canonical, martyna1994constant}

After equilibration at 700 K, the systems were gradually cooled to 200 K to observe any transition to an ordered state. Cooling was simulated in the NPT ensemble at 1 atm, with temperature decrements of 2 K and 2 ns of simulation per step, totaling 500 ns and achieving a cooling rate of 1 K/ns. At the end of each temperature step, the order parameter (P2) was calculated from the structure with molecular directions defined by the principal axes of inertia. The transition temperature (T\textsubscript{ni}) was identified as the lowest temperature where P2 crossed the threshold value of 0.2. T\textsubscript{ni} values from the two replicates were averaged. To further confirm the presence of the liquid crystal phase, for selected molecules, a reverse simulation was performed, where the system was gradually heated from an ordered to a disordered state to check the observed phase transition temperatures from the cooling process.

\section{Experimental UV-Vis Measurements}
The polarization-dependent UV-Vis transmission and reflection spectra were recorded using a Cary 7000 spectrophotometer. Measurements were conducted over a wavelength range of 250 nm to 750 nm, with data collected in 1 nm increments and an averaging time of 0.5 sec per measurement. The spectral bandwidth of the incident light was set to 4 nm on the monochromator, further refined by a pair of 3-degree vertical slits and a 1-degree horizontal slit, producing a square beam approximately 4 mm in lateral length. Baseline measurements for 100\% transmission and zero were obtained at a detector angle of 180 degrees, using air as the reference. Samples were mounted vertically and aligned with the detector at an incidence angle of 6 degrees using white light. Subsequent data collection was also conducted at a 6-degree angle of incidence, with the auto-polarizer set to 0 and 90 degrees for S and P polarized light, respectively.

\section{Fitness Function Design}
\label{appendix:fitness function}

\textbf{Fitness function 1} consists of three components: the transparency score (\(S_{\text{transparency}}\)), the average refractive index score (\(S_{n_{\text{avg}}}\)), and the anisotropic refractive index difference score (\(S_{n_{\text{diff}}}\)). The selection prioritizes molecules with high refractive indices while maintaining optical transparency. A molecule qualifies for scoring and ranking if it meets the following criteria:  

1. The maximum absorption wavelength (\(\lambda_{\text{max}}\)) is below 460 nm:  
   \[
   \lambda_{\text{max}} < 460\,\text{nm}
   \]

2. The ratio of absorbance at \(\lambda_{\text{max}}\) to absorbance at 460 nm is greater than 1:  
   \[
   \frac{A(\lambda_{\text{max}})}{A(460\,\text{nm})} \geq 1
   \]
   
3. The anisotropic refractive index difference, defined as \(\Delta n\), is above 0.3:  
   \[
   \Delta n = n_e - n_o \geq 0.3
   \]

The transparency score is defined as a piecewise function:  
\[
S_{\text{transparency}} =
\begin{cases}  
\log \left( \frac{A(460\,\text{nm})}{A(\lambda_{\text{max}})} \right), & \text{if } \frac{A(\lambda_{\text{max}})}{A(460\,\text{nm})} < 100 \\  
8, & \text{if } 100 \leq \frac{A(\lambda_{\text{max}})}{A(460\,\text{nm})} < 1000 \\  
10, & \text{if } \frac{A(\lambda_{\text{max}})}{A(460\,\text{nm})} \geq 1000  
\end{cases}
\]

The average refractive index score is defined as:  
\[
S_{n_{\text{avg}}} = n_{\text{avg}} \times 10
\]

The anisotropic refractive index difference score (\(S_{n_{\text{diff}}}\)) is defined as:  
\[
S_{n_{\text{diff}}} = \Delta n \times 20 = (n_e - n_o) \times 20
\]

Finally, the overall Fitness Function 1 score (\(S_{\text{fitness1}}\)) is computed as the sum of the three individual scores:  
\[
S_{\text{fitness1}} = S_{\text{transparency}} + S_{n_{\text{diff}}} + S_{n_{\text{avg}}}
\]

\textbf{Fitness function 2} consists of four components: the transparency score (\(S_{\text{transparency}}\)), the average refractive index score (\(S_{n_{\text{avg}}}\)), the anisotropic refractive index difference score (\(S_{n_{\text{diff}}}\)), and the maximum wavelength score (\(S_{\lambda_{\text{max}}}\)). This function prioritizes molecules with high transparency while maintaining a high refractive index. A molecule qualifies for scoring and ranking if it meets the following criteria:  

1. The average refractive index is above 1.5:  
   \[
   n_{\text{avg}} \geq 1.5
   \]
2. The anisotropic refractive index difference, defined as \(\Delta n\), is above 0.3:  
   \[
   \Delta n = n_e - n_o \geq 0.3
   \]

The maximum wavelength score is defined as:

\[
S_{\lambda_{\text{max}}} = \frac{460 - \lambda_{\text{max}}}{5}
\]

The transparency score is defined as a piecewise function:

\[
S_{\text{transparency}} = 
\begin{cases} 
\log\left( \frac{A(\lambda 460\text{nm})}{A(\lambda \text{max})} \right) \times 10 + S_{\lambda_{\text{max}}}, & \text{if } \frac{A(\lambda \text{max})}{A(\lambda 460\text{nm})} \leq 10 \\
30 + S_{\lambda_{\text{max}}}, & \text{if } 10 \leq \frac{A(\lambda \text{max})}{A(\lambda 460\text{nm})} \leq 100 \\
80 + S_{\lambda_{\text{max}}}, & \text{if } 100 \leq \frac{A(\lambda \text{max})}{A(\lambda 460\text{nm})} \leq 1000 \\
100 + S_{\lambda_{\text{max}}}, & \text{if } \frac{A(\lambda \text{max})}{A(\lambda 460\text{nm})} \geq 1000 
\end{cases}
\]

The anisotropic refractive index difference score is defined as:

\[
S_{n_{\text{diff}}} = \Delta n \times 20
\]

The average refractive index score is defined as:

\[
S_{n_{\text{avg}}} = n_{\text{avg}} \times 100
\]

Finally, the overall Fitness Function 2 score (\(S_{\text{fitness2}}\)) is computed as the sum of the four individual scores:  
\[
S_{\text{fitness2}} = S_{\text{transparency}} + S_{n_{\text{diff}}} + S_{n_{\text{avg}}} + S_{\lambda_{\text{max}}}\
\]

\clearpage
\newpage
\section{Tables}
\setcounter{table}{0}
\renewcommand{\thetable}{S\arabic{table}}

\begin{table}[!h]
    \centering
    \caption{The nematic transition temperature (T\textsubscript{ni}) of the five commercial reactive mesogens: HCM-008, HCM-009, HCM-020, HCM-021, and HCM-083.}
    \label{tab:benchmark-tni}
    \begin{tabular}{cccccc}
    \toprule
    Sample & T\textsubscript{ni,\ avg} (K) & T\textsubscript{ni,\ cool$_{1}$} (K) & T\textsubscript{ni,\ cool$_{2}$} (K) & T\textsubscript{ni,\ heat$_{1}$} (K) & T\textsubscript{ni,\ heat$_{2}$} (K) \\
    \midrule
    HCM-008 & 515 & 495 & 469 & 553 & 543 \\
    HCM-009 & 467  & 479 & 383 & 503 & 503 \\
    HCM-020 & 410 & low & 357 & 433 & 439 \\
    HCM-021 & 410 & low & low & 411 & 409 \\
    HCM-083 & 499 & 493 & 507 & 495 & 501 \\
    \bottomrule
    \end{tabular}
\end{table}

\begin{table}[!h]
    \centering
    \caption{The maximum absorption wavelengths (in nm) of the average spectra for the five reactive mesogens calculated via dimer pipeline (Dimer-xTB) and MD simulations (Dimer-MD), compared with experimental spectra (Experiment). The MD spectra were averaged from 200 dimers in the nematic phase, while the dimer pipeline spectra were averaged from the 8 lowest energy dimers. When compared to experiments, the mean absolute deviation (MAE) of the dimer-xTB approach is 9.4 nm and that of the Dimer-MD approach is 6.3 nm.}
    \label{tab:uv-table}
    \begin{tabular}{ccccccc}
    \toprule
    Approach & HCM-008 & HCM-009 & HCM-020 & HCM-021 & HCM-083 \\
    \midrule
    Dimer-xTB & 274 & 273 & 268 & 254 & 259 \\
    Dimer-MD & 260 & 262 & 254 & 250 & 252 \\
    Experiment & 257 & 257 & 267 & 260 & 252 \\
    \bottomrule
    \end{tabular}
\end{table}

\renewcommand{\arraystretch}{1.2}
\begin{table}[!h]
    \centering
    \caption{Density, polarizability ($\alpha_{\text{avg}}$), and isotropic refractive index ($n_{\text{avg}}$) values of the dimers and their constituent monomers ($m_{\text{1}}$, $m_{\text{2}}$) for four reactive mesogens and their various conformations.}
    \label{tab:dimer-RI-table}
    {\scriptsize
    \begin{tabular}{c@{\hskip 5pt}c@{\hskip 6pt}c@{\hskip 5pt}c@{\hskip 5pt}c@{\hskip 5pt}c@{\hskip 5pt}c@{\hskip 5pt}c@{\hskip 6pt}c@{\hskip 5pt}c@{\hskip 5pt}c@{\hskip 5pt}c@{\hskip 5pt}c@{\hskip 5pt}c@{\hskip 6pt}c@{\hskip 5pt}c@{\hskip 5pt}c@{\hskip 5pt}c@{\hskip 5pt}c}
    \toprule
    \multirow{2}{*}{Sample} & \multicolumn{5}{c}{Density (g/cm\textsuperscript{3})} & & \multicolumn{5}{c}{$\alpha_{\text{avg}}$} & & \multicolumn{5}{c}{$n_{\text{avg}}$} \\ 
    
    \cmidrule{3-7} \cmidrule{9-13} \cmidrule{15-19} 
    
    & & $m_{\text{1}}$  & $m_{\text{2}}$  & $m_{\text{1}} + m_{\text{2}}$ & dimer & $\Delta$\% & & $m_{\text{1}}$  & $m_{\text{2}}$  & $m_{\text{1}} + m_{\text{2}}$ & dimer & $\Delta$\% & & $m_{\text{1}}$ & $m_{\text{2}}$ & $m_{\text{1}} + m_{\text{2}}$ & dimer & $\Delta$\% \\ 
    
    \midrule
    HCM020-1 & & 1.1732  & 1.1735  & 1.1733  & 1.1747  & 0.12  & & 283.19  & 282.87  & 566.06  & 570.75  & 0.82  & & 1.5439 & 1.5433 & 1.5436 & 1.5499 & 0.40\\
    HCM020-2 & & 1.1726  & 1.1733  & 1.1729  & 1.1739  & 0.08  & & 283.50  & 283.21  & 566.71  & 572.33  & 0.98  & & 1.5443 & 1.5440 & 1.5442 & 1.5513 & 0.46\\
    HCM020-3 & & 1.1728  & 1.1735  & 1.1732  & 1.1728  & 0.03  & & 283.28  & 282.93  & 566.21  & 564.38  & 0.32  & & 1.5435 & 1.5439 & 1.5437 & 1.5414 & 0.15\\
    HCM020-4 & & 1.1728  & 1.1728  & 1.1728  & 1.1728  & 0.00  & & 283.18  & 283.29  & 566.46  & 563.78  & 0.48  & & 1.5440 & 1.5437 & 1.5438 & 1.5407 & 0.20\\
    HCM021-1 & & 1.1540  & 1.1582  & 1.1561  & 1.1720  & 1.36  & & 282.31  & 281.66  & 563.97  & 561.06  & 0.52  & & 1.5233 & 1.5241 & 1.5237 & 1.5291 & 0.35\\
    HCM021-2 & & 1.1553  & 1.1560  & 1.1557  & 1.1596  & 0.33  & & 283.81  & 284.13  & 567.94  & 565.55  & 0.42  & & 1.5273 & 1.5284 & 1.5279 & 1.5274 & 0.04\\
    HCM021-3 & & 1.1573  & 1.1567  & 1.1570  & 1.1604  & 0.29  & & 281.61  & 282.21  & 563.81  & 571.26  & 1.30  & & 1.5235 & 1.5245 & 1.5240 & 1.5342 & 0.66\\
    HCM021-4 & & 1.1578  & 1.1578  & 1.1578  & 1.1580  & 0.02  & & 281.58  & 282.00  & 563.58  & 569.08  & 0.97  & & 1.5237 & 1.5247 & 1.5242 & 1.5304 & 0.41\\
    HCM021-5 & & 1.1571  & 1.1573  & 1.1572  & 1.1573  & 0.01  & & 281.60  & 282.13  & 563.74  & 561.90  & 0.33  & & 1.5234 & 1.5247 & 1.5241 & 1.5221 & 0.13\\
    HCM083-1 & & 1.1592  & 1.1592  & 1.1592  & 1.1589  & 0.02  & & 282.06  & 282.05  & 564.11  & 563.04  & 0.19  & & 1.5256 & 1.5255 & 1.5256 & 1.5242 & 0.09\\
    HCM083-2 & & 1.1591  & 1.1592  & 1.1592  & 1.1590  & 0.01  & & 282.05  & 282.05  & 564.11  & 564.42  & 0.06  & & 1.5255 & 1.5255 & 1.5255 & 1.5258 & 0.02\\
    HCM083-3 & & 1.1592  & 1.1592  & 1.1592  & 1.1588  & 0.03  & & 282.05  & 564.10  & 564.10  & 561.46  & 0.47  & & 1.5255 & 1.5255 & 1.5255 & 1.5224 & 0.20\\
    HCM083-4 & & 1.1591  & 1.1592  & 1.1592  & 1.1588  & 0.03  & & 282.06  & 564.10  & 564.12  & 564.03  & 0.02  & & 1.5255 & 1.5256 & 1.5256 & 1.5253 & 0.02\\
    HCM083-5 & & 1.1592  & 1.1592  & 1.1592  & 1.1590  & 0.02  & & 282.05  & 564.12  & 564.10  & 562.13  & 0.35  & & 1.5255 & 1.5255 & 1.5255 & 1.5233 & 0.15\\
    HCM009-1 & & 1.1620  & 1.1567  & 1.1593  & 1.1545  & 0.42  & & 489.63  & 491.01  & 980.63  & 979.96  & 0.07  & & 1.5449 & 1.5437 & 1.5443 & 1.5411 & 0.21\\
    \bottomrule
    \end{tabular}
    }
    
\end{table}
\renewcommand{\arraystretch}{1}

\begin{table}[!h]
    \centering
    \caption{The nematic transition temperature (T\textsubscript{ni}) of ten candidate liquid crystal structures. Average and individual results are reported from MD cooling process for different random initializations of the systems.}
    \label{tab:candidates-table}
    \begin{tabular}{cccc}
    \toprule
    Sample & T\textsubscript{ni,\ avg} (K) & T\textsubscript{ni,\ cool$_{1}$} (K) & T\textsubscript{ni,\ cool$_{2}$} (K) \\
    \midrule
    Candidate1 & <200 & <200 & <200 \\
    Candidate2 & 504 & 505 & 503 \\
    Candidate3 & 623 & 619 & 617 \\
    Candidate4 & 642 & 645 & 639 \\
    Candidate5 & 666 & 675 & 657 \\
    Candidate6 & 671 & 675 & 667 \\
    Candidate7 & 699 & 699 & >700 \\
    Candidate8 & >700 & >700 & >700 \\
    Candidate9 & >700 & >700 & >700 \\
    Candidate10 & >700 & >700 & >700 \\
    \bottomrule
    \end{tabular}
\end{table}

\clearpage
\newpage
\section{Figures}

\begin{figure*}[!htb]
     \centering
     \includegraphics[width=\textwidth]{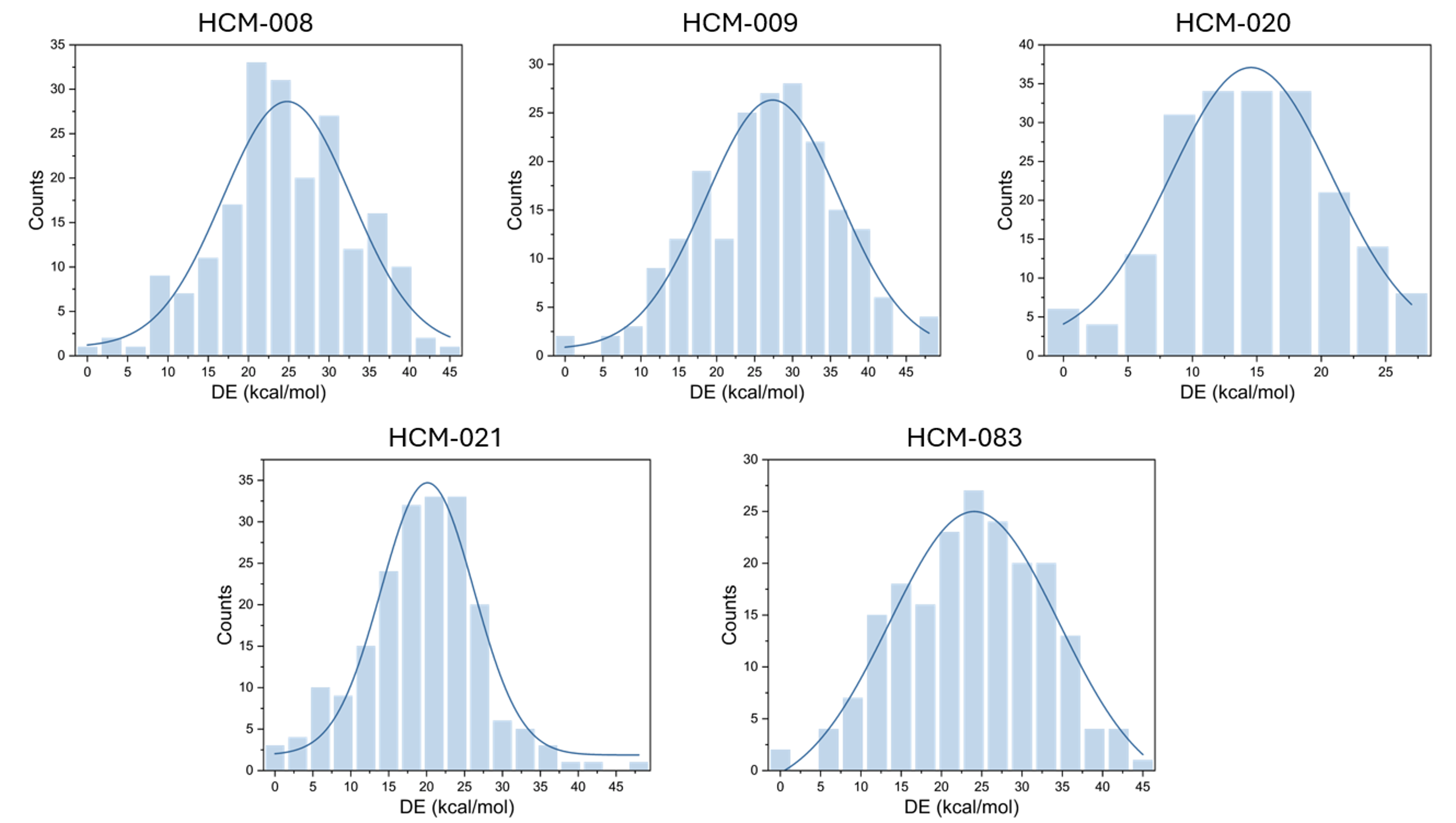}
     \caption{Energy distributions of 200 MD-extracted dimers from five commercial reactive mesogens in the nematic phase calculated using the GFN2-xTB method.}
     \label{fig:distribution}
\end{figure*}

\begin{figure*}[!htb]
     \centering
     \includegraphics[width=\textwidth]{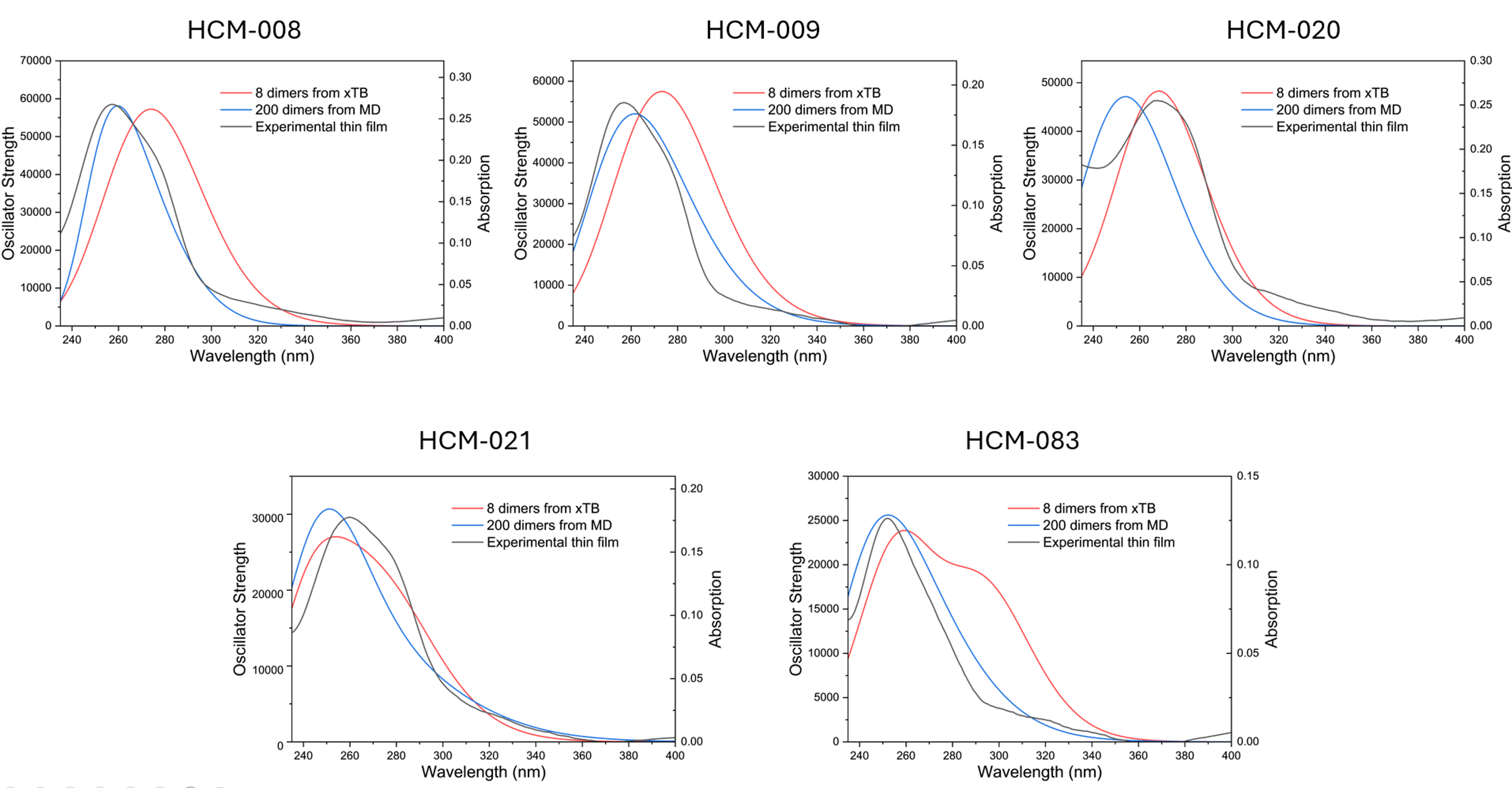}
     \caption{Experimental thin film absorption spectra of the five reactive mesogens. Each plot also contains average spectra of 200 MD-extracted dimers in nematic phase and average spectra for 8 dimers extracted from dimer generation pipeline.}
     \label{fig:spectra-all}
\end{figure*}

\begin{figure*}[!htb]
     \centering
     \includegraphics[width=0.98\textwidth]{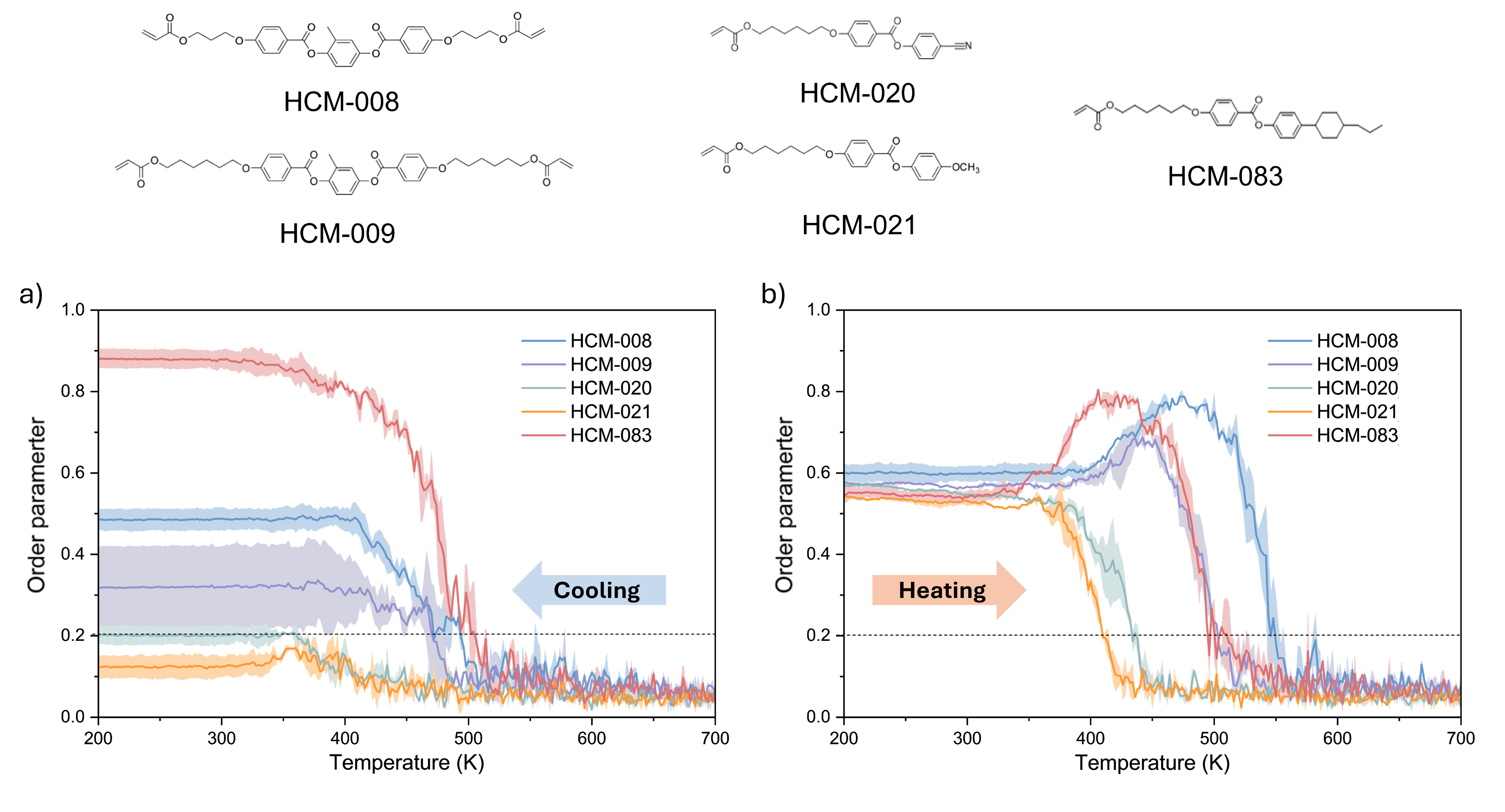}
     \caption{Order parameter as a function of temperature for HCM-008, HCM-009, HCM-020, HCM-021, and HCM-083 during the MD cooling and heating processes.}
     \label{fig:Benchmark-p2}
\end{figure*}

\begin{figure*}[!htb]
     \centering
     \includegraphics[width=0.98\textwidth]{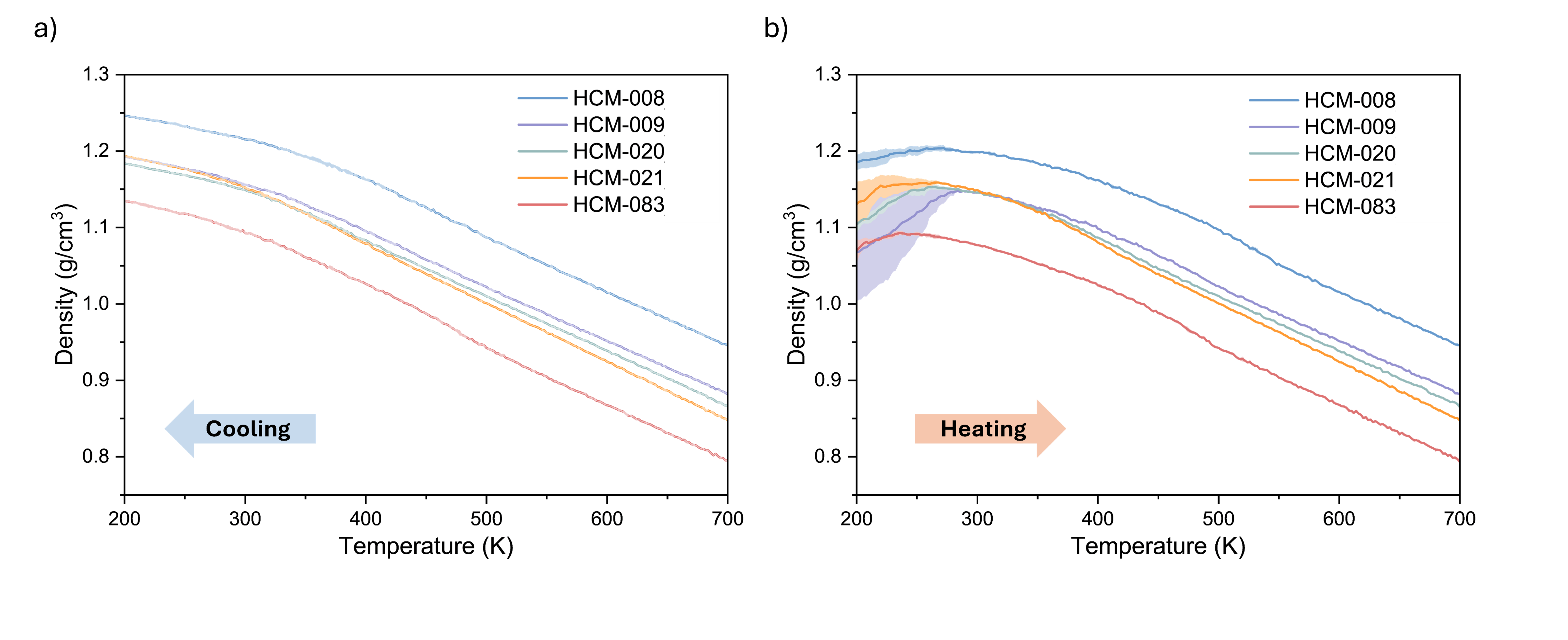}
     \caption{Density as a function of temperature for HCM-008, HCM-009, HCM-020, HCM-021, and HCM-083 during the MD cooling and heating processes.}
     \label{fig:Density-Temperature}
\end{figure*}

\begin{figure*}[htbp]
     \centering
     \includegraphics[width=0.75\textwidth]{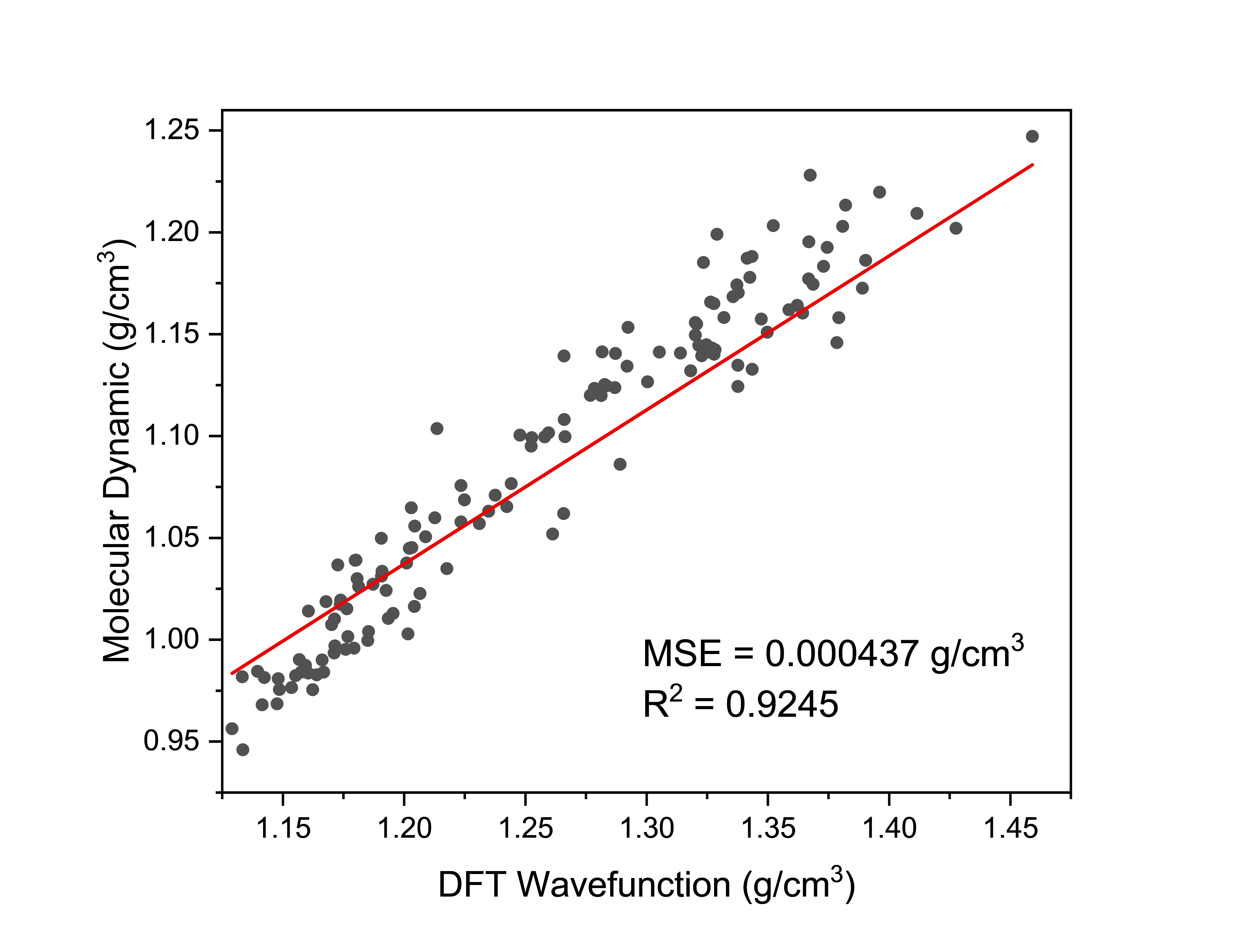}
     \caption{The density calculated by the van der Waals molecular volume via DFT wavefunctions compared to the cell density of the molecules at 298 K during MD cooling process from 700 K to 200 K.}
     \label{fig:density-benchmark}
\end{figure*}

\begin{figure*}[!htb]
     \centering
     \includegraphics[width=\textwidth]{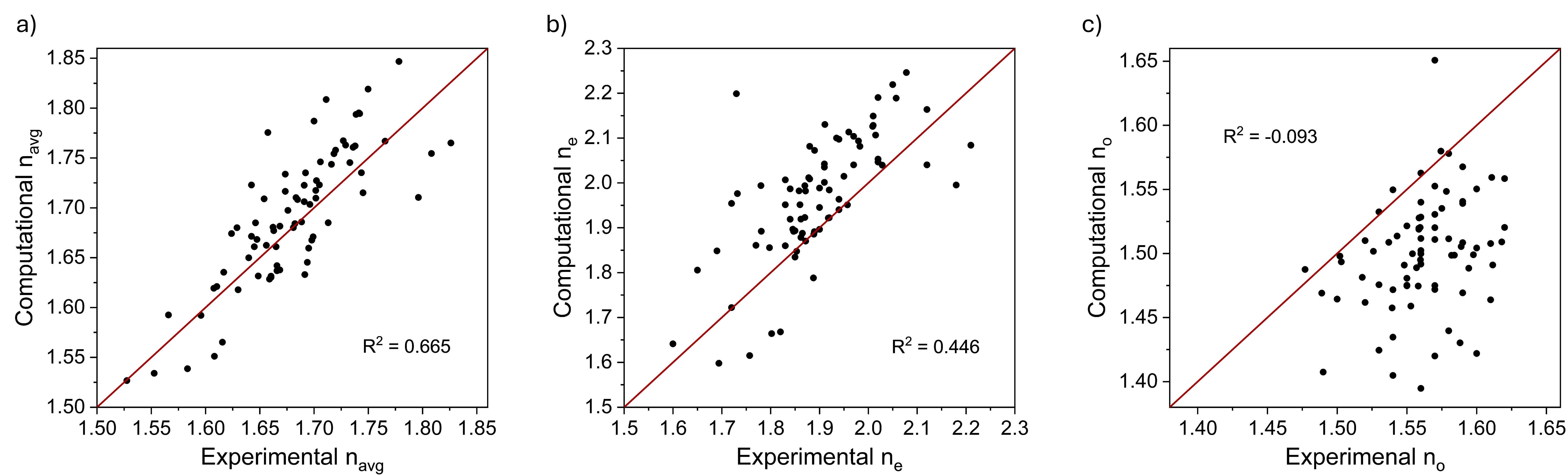}
     \caption{(a) $n_{avg}$, (b) $n_{e}$, and c) $n_{o}$ refractive index comparisons of the 77 literature-reported liquid crystal small molecules obtained from DFT calculations using the M06-HF functional along with the ZPOL basis set and molecular volume calculated density approach.}
     \label{fig:RI-all}
\end{figure*}

\begin{figure*}[htbp]
     \centering
     \includegraphics[width=0.98\textwidth]{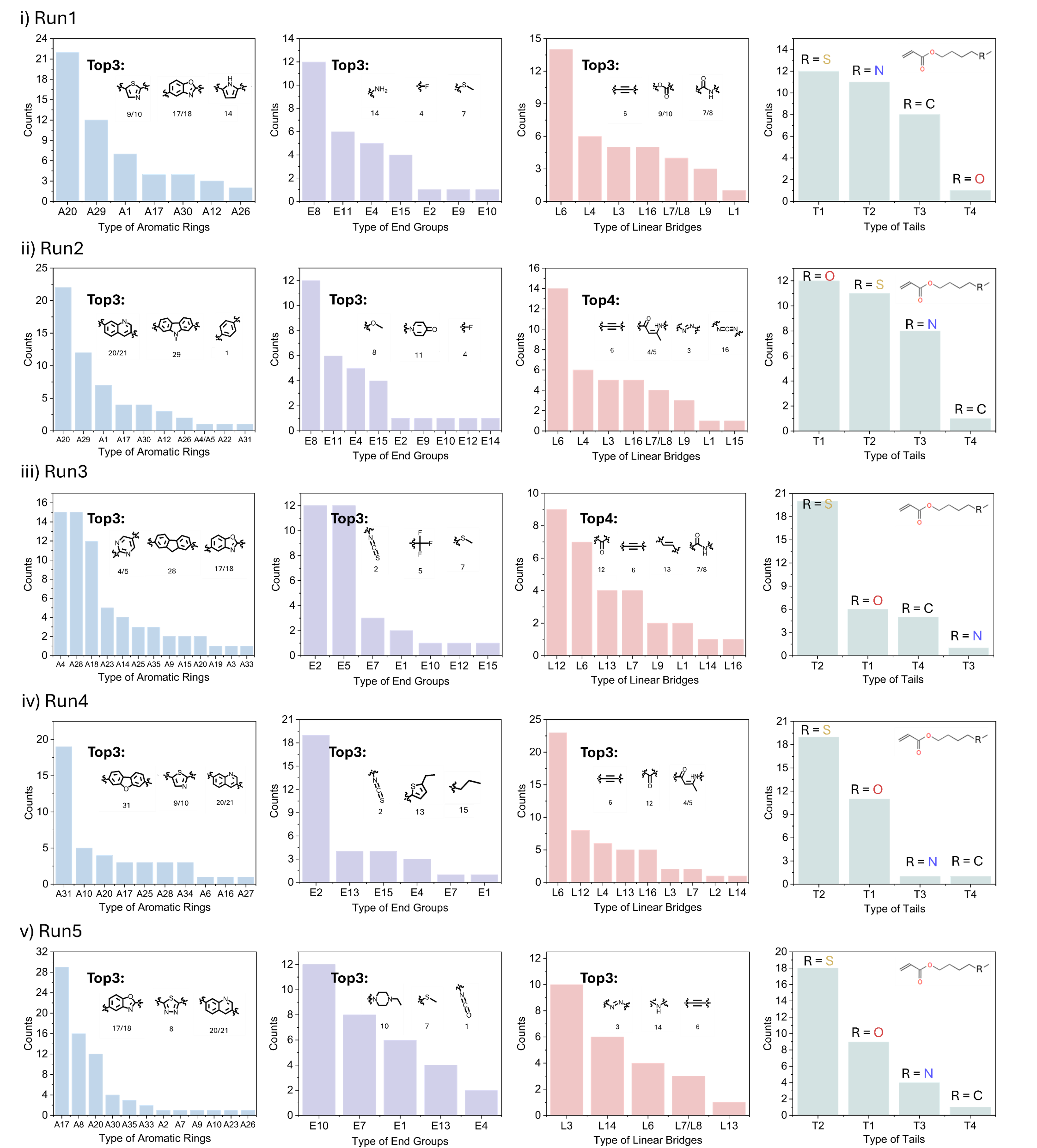}
     \caption{Frequency of the building blocks for Fitness function 1 and the top candidates for each component in monomer structure generation.}
     \label{fig:Freq-fit1-all}
\end{figure*}

\begin{figure*}[htbp]
     \centering
     \includegraphics[width=0.98\textwidth]{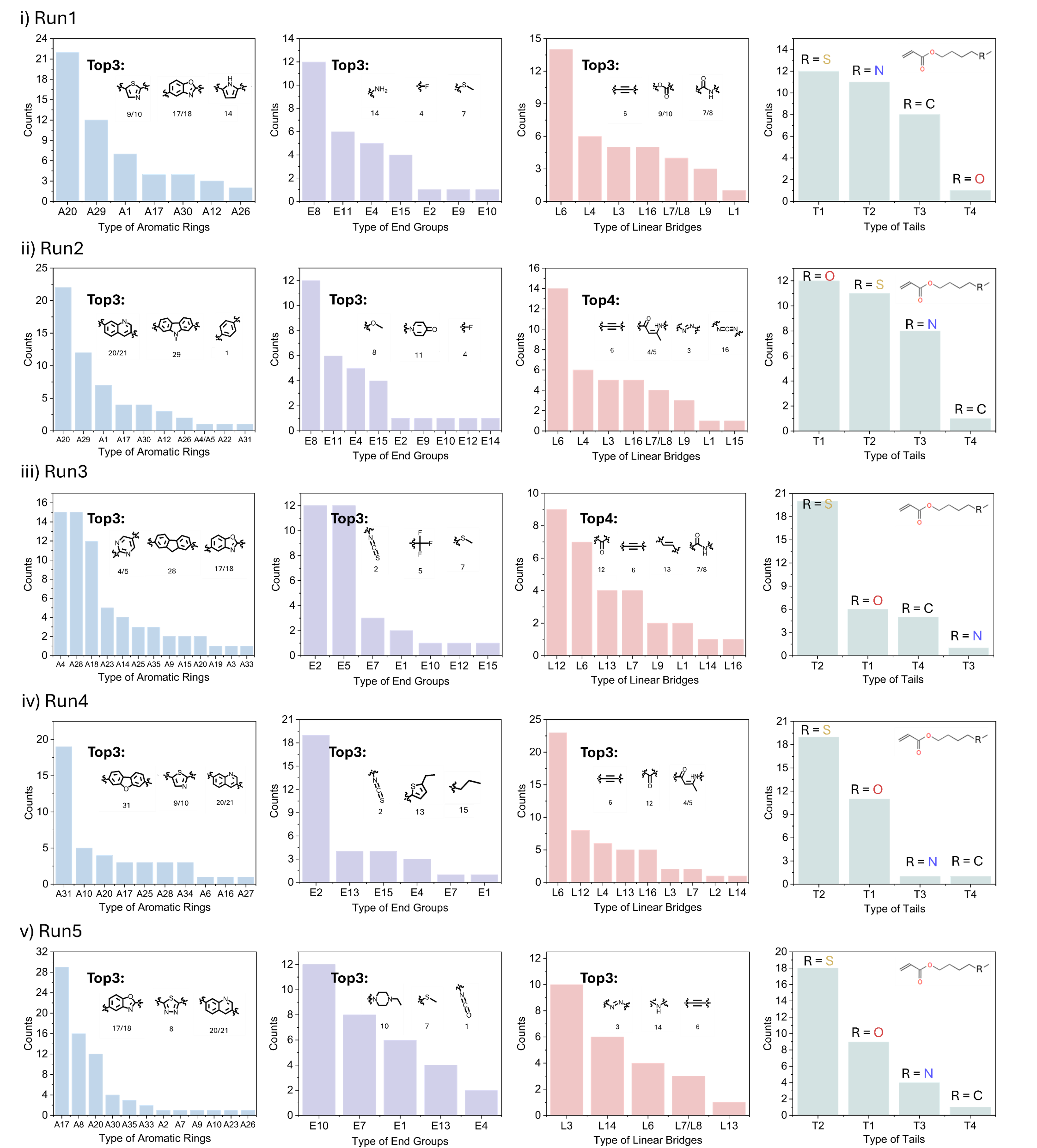}
     \caption{Frequency of the building blocks for Fitness function 2 and the top candidates for each component in monomer structure generation.}
     \label{fig:Freq-fit2-all}
\end{figure*}

\begin{figure*}[!htbp]
     \centering
     \includegraphics[width=0.9\textwidth]{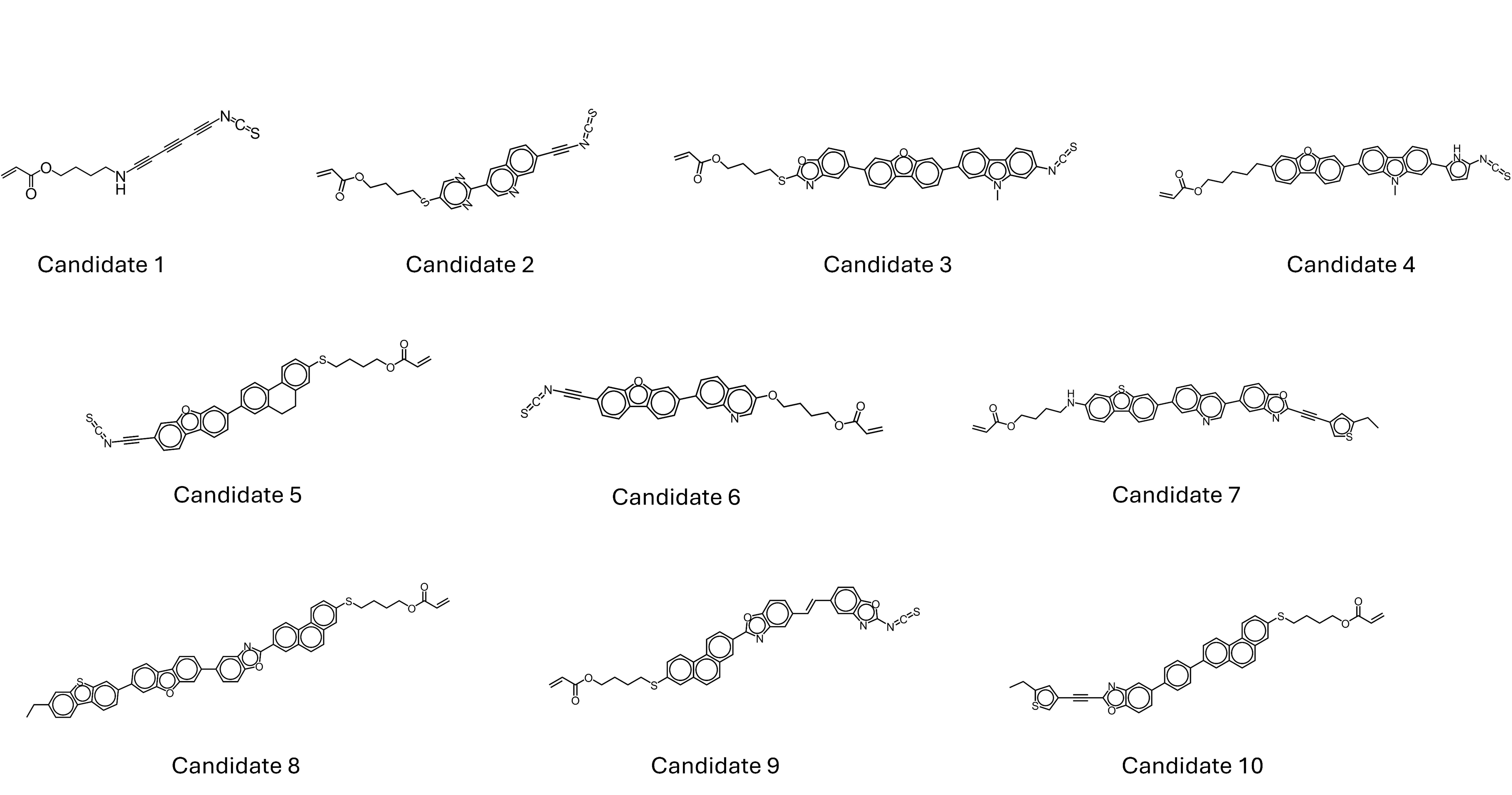}
     \includegraphics[width=0.75\textwidth]{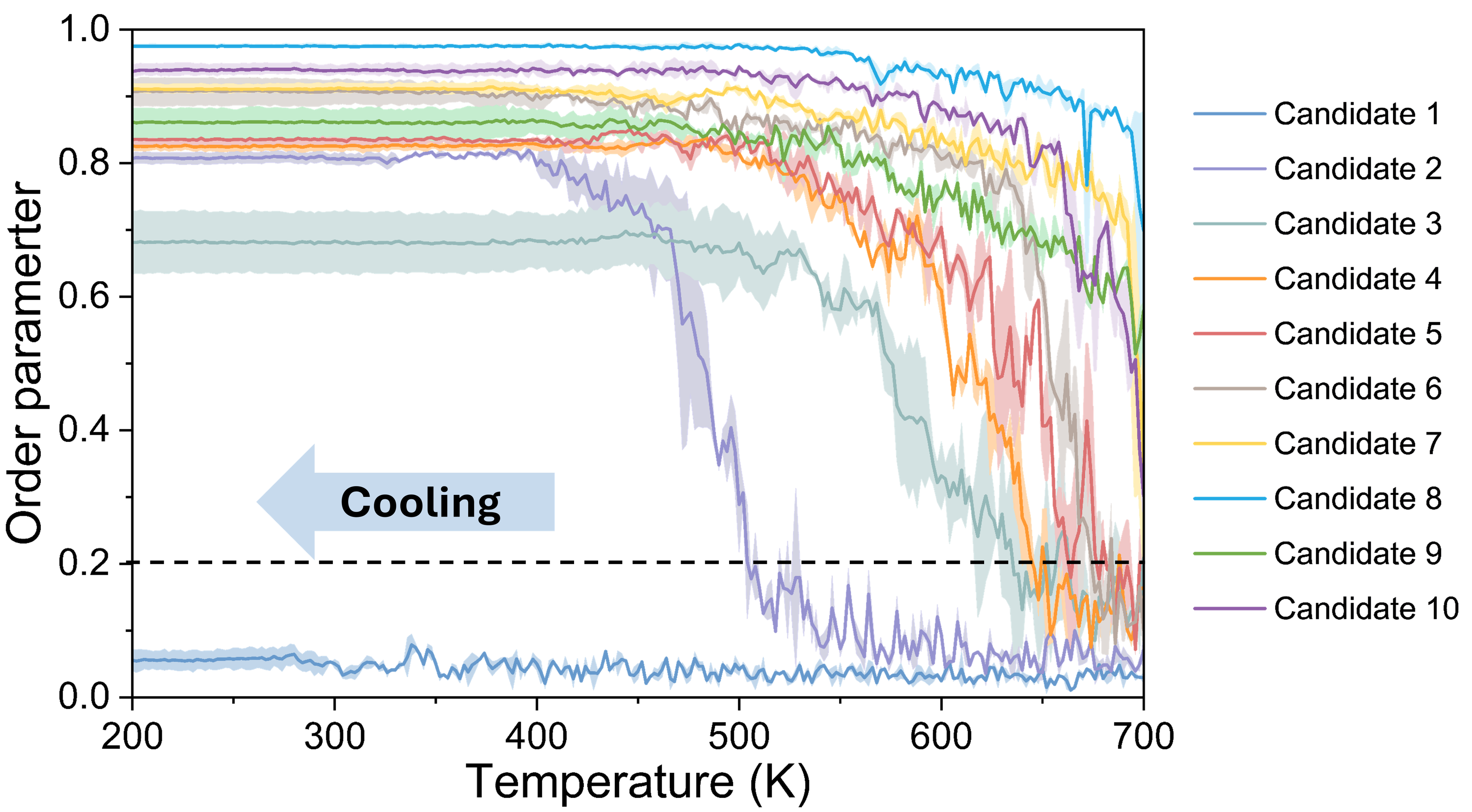}
     \caption{Order parameter as a function of temperature for all ten candidates during MD cooling process. For each candidate, the figure shows the average of the order parameter (darker colors), as well as the maximum and minimum values at the given temperature (lighter colors) for two MD cooling runs for different random initializations of the starting system. Overall, both runs are consistent for all candidates.}
     \label{fig:candidates-P2}
\end{figure*}

\begin{figure*}[!htbp]
     \centering
     \includegraphics[width=0.7\textwidth]{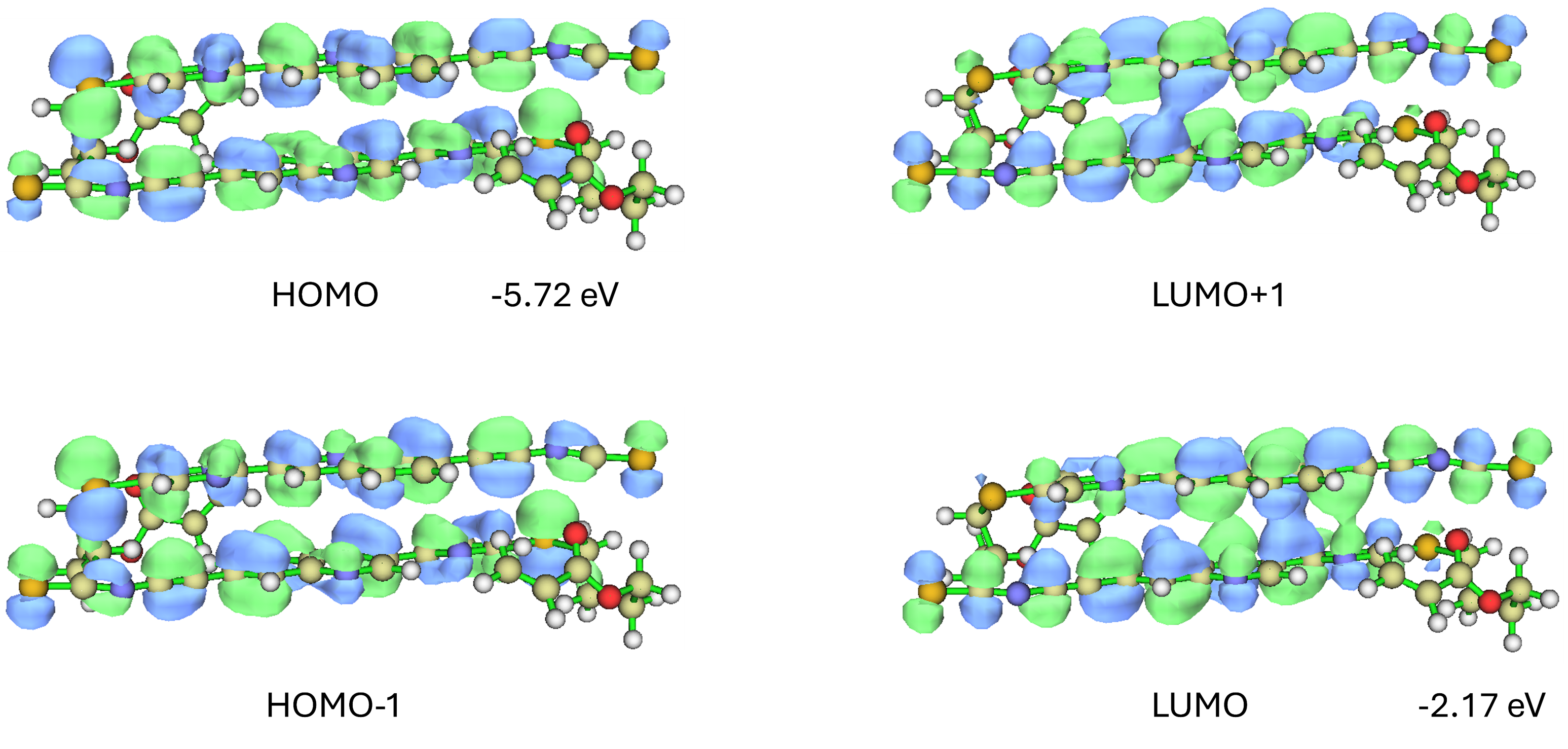}
     \caption{Molecular orbitals of Candidate 2, with the green and blue isosurfaces representing positive and negative values, respectively.}
     \label{fig:candidate2-orbitals}
\end{figure*}

\clearpage
\newpage
\subsection{Building Blocks List}

\begin{figure*}[!htb]
     \centering
     \includegraphics[width=0.9\textwidth]{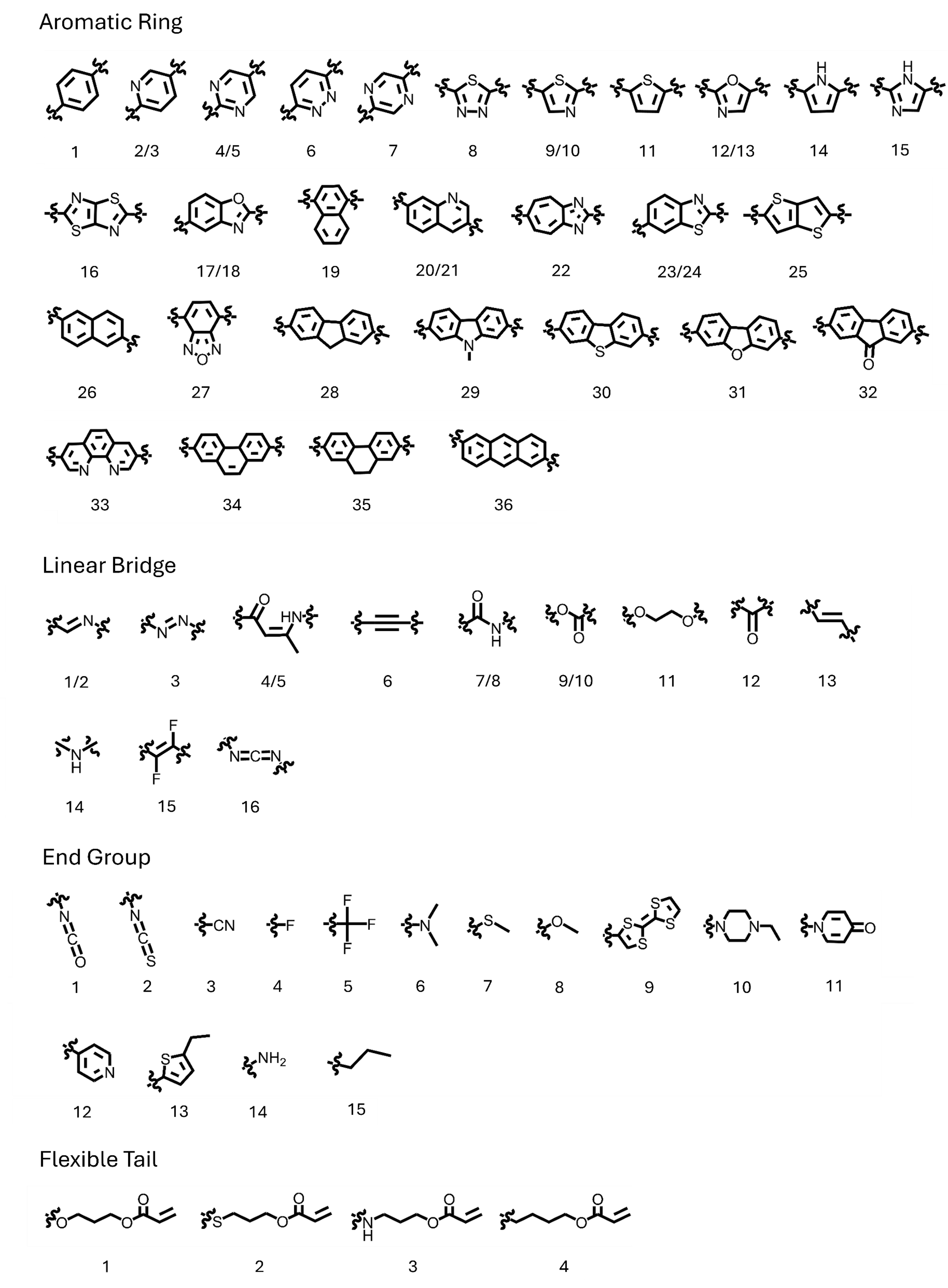}
     \caption{Building blocks list of the four components constituting the reactive mesogen structure: aromatic rings, linear bridges, tails, and end groups. Several of aromatic rings and linear bridges are represented with two indices, meaning that there in an extra choice on the connectivity direction.}
     \label{fig:BB-database}
\end{figure*}

\end{document}